\documentclass[preprintnumbers]{revtex4}

\usepackage{graphicx}
\newcommand{\mh}{\mbox{$m_{\rm{h}}$}}
\newcommand{\mH}{\mbox{$m_{\rm{H}}$}}
\newcommand{\mw}{\mbox{$m_{\rm{W}}$}}
\newcommand{\mt}{\mbox{$m_{\rm{t}}$}}

\def \etal{{\it et al.}}

\begin{document}
% You should use BibTeX and revtex.bst for references
\bibliographystyle{revtex}

% number  and your conference paper identification number on the
% title page in preprint mode. Multiple \preprint commands are allowed.
\preprint{Fermilab-Conf-01/307-T}
\preprint{MADPH-01-1243} 

\vspace*{.1in}

\title{E1 Working Group Summary: Neutrino Factories and Muon Colliders}
% Optional argument for running titles on pages
%\title[]{}



\author{T.~Adams} 
\affiliation{Florida State University, Tallahassee, FL 32306, USA} 
\author{C.~Albright}
\author{V.~Balbekov} 
\author{G.~Barenboim} 
\author{D.~A.~Harris} 
\email[]{dharris@fnal.gov}
\author{W.~Chou}
\author{F.~DeJongh}
\author{S.~Geer}
\author{C.~Johnstone}
\author{N.~Mokhov}
\author{J.~Morfin}
\author{D.~Neuffer}
\author{R.~Raja}
\author{A.~Romanino}
\author{P.~Shanahan}
\author{P.~Spentzouris}
\author{J.~Yu}
\affiliation{Fermi National Accelerator Laboratory, Batavia, IL 60510, USA}
\author{V.~Barger}
\email[]{barger@oriole.physics.wisc.edu}
\author{D.~Marfatia} 
\author{T.~Han} 
\affiliation{University of Wisconsin, Madison, WI 53706, USA}
\author{M.~Aoki}
\author{Y.~Kuno}
\email[]{kuno@phys.sci.osaka-u.ac.jp}
\author{A.~Sato} 
\affiliation{Osaka University, Toyonaka, Osaka 560-0043, Japan} 
\author{K.~Ichikawa} 
\author{T.~Nakaya}
\affiliation{Kyoto University, Kyoto 606-8502, Japan} 
\author{S.~Machida}
\author{K.~Nagamine}
\author{K.~Yoshimura}
\affiliation{High Energy Accelerator Research Organization (KEK), 
Tsukuba, Ibaraki 305-0801, Japan} 
\author{R.~D.~Ball}
\affiliation{CERN, European Organization for Nuclear Research, CH1211
Geneva 23 Switzerland and University of Edinburgh, 
Edinburgh EH9 3JZ, United Kingdom}
\author{M.~Campanelli} 
\affiliation{University of Geneva, CH-1211 Geneva 4, Switzerland and 
Institut fur Teilchenphysik, ETHZ, CH-8093, Zurich, Switzerland} 
\author{D.~Casper}
\author{W.~Molzon}
\author{H.~Sobel}
\affiliation{University of California, Irvine, CA 92697, USA} 
\author{D.~B.~Cline} 
\affiliation{University of California, Los Angeles, CA 90095, USA} 
\author{P.~Cushman} 
\affiliation{University of Minnesota, Minneapolis, MN 55455, USA}
\author{M.~Diwan}
\author{S.~Kahn}
%\author{B.~King}
%\author{W.~Marciano}
\author{W.~Morse} 
\author{R.~Palmer}
\author{Z.~Parsa}
\author{T.~Roser}
\affiliation{Brookhaven National Laboratory, Upton, NY 11973, USA} 
\author{B.~T.~Fleming} 
\author{J.~A.~Formaggio} 
\affiliation{Columbia University, New York, NY 10027, USA} 
\author{A.~Garren} 
\affiliation{Lawrence Berkeley National Laboratory, Berkeley, CA 94720, USA}
\author{M.~B.~Gavela} 
\affiliation{Univ. Aut\'{o}noma de Madrid, Madrid, 28049, Spain}  
\author{M.~C.~Gonzalez-Garcia} 
\affiliation{CERN, European Organization for Nuclear Research, CH1211
Geneva 23 Switzerland and Universitat de Val\`{e}ncia - C.S.I.C. Val\`{e}ncia, 46071 Spain} 
\author{G.~Hanson}
\author{M.~Berger}
\affiliation{Indiana University, Bloomington, IN, 47405 USA} 
\author{B.~Kayser}
\affiliation{National Science Foundation, Arlington, VA 22230, USA}
\author{C.~K.~Jung}
\author{R.~Shrock}
\author{C.~McGrew} 
\author{I.~Mocioiu}
\affiliation{State University of New York, Stonybrook, NY 11794, USA} 
\author{M.~Lindner}
\affiliation{Technische Universitat Munchen 85748 Garching, Germany}
\author{K.~McDonald} 
\affiliation{Princeton University, Princeton, NJ 08544, USA} 
\author{K.~S.~McFarland} 
\affiliation{University of Rochester, Rochester, NY 14627, USA} 
\author{P.~Nienaber}
\affiliation{College of the Holy Cross, Worcester, MA 01610, USA}
\author{F.~Olness} 
\affiliation{Southern Methodist University, Dallas, TX 75275, USA} 
\author{B.~Pope} 
\affiliation{Michigan State University, East Lansing, MI 48824, USA}
\author{S.~Rigolin} 
\affiliation{University of Michigan, Ann Arbor, MI 48109, USA}
\author{L.~Roberts}
\affiliation{Boston University, Boston, MA 02215, USA}
\author{H.~Schellman}
\affiliation{Northwestern University, Evanston, IL 60208, USA}
\author{M.~Shiozawa} 
\affiliation{Institute of Cosmic Ray Research, University of Tokyo,
Tokyo 113-0033, Japan} 
\author{L.~Wai}
\affiliation{Stanford University, Stanford, CA 94305, USA} 
\author{Y.~F.~Wang}
\affiliation{Institute of High Energy Physics, Beijing, 100039 China} 
\author{K.~Whisnant} 
\affiliation{Iowa State University, Ames, IA 50011, USA} 
\author{M.~Zeller} 
\email[]{michael.zeller@yale.edu}
\affiliation{Yale University, New Haven CT 06520, USA} 
%\date{\today}

\begin{abstract}
We are in the middle of a time of exciting discovery, namely that neutrinos
have mass and oscillate.  In order to take the next steps to understand
this potential window onto what well might be the mechanism that 
links the quarks and leptons, we need
both new neutrino beams and new detectors.  The new beamlines
can and should also provide new laboratories for doing charged
lepton flavor physics, and the new detectors can and should also provide
laboratories for doing other physics like proton decay, supernovae
searches, etc.  The new neutrino beams serve as milestones along the 
way to a muon collider, which can answer questions in  
yet another sector of particle physics, namely the Higgs sector 
or ultimately the energy frontier.  
In this report we discuss the current status of 
neutrino oscillation physics, what other oscillation measurements
are needed to fully explore the phenomenon, and finally, 
what other new physics can be explored as a result of building 
of these facilities.  
\end{abstract}
% \pacs{}

%\maketitle must follow title, authors, abstract and \pacs
\maketitle

\section{Introduction} 

The experimental study 
of the fundamental properties of neutrinos is one
of the most challenging in particle physics, but it has also produced
some of the most exciting and revolutionary physics insights, both on
microscopic and astrophysical scales~\cite{exp}.  
The discovery of the
electron neutrino emitted in beta decay in the 1950s confirmed the
hypothesis in the early 1930s by Pauli and Fermi of the
existence of the electron neutrino.  Shortly thereafter its left-handed
nature was confirmed,  the muon neutrino was discovered, and
oscillations among three neutrino flavors were predicted.  Neutrino
neutral currents were pivotal in the confirmation of the Standard Model
in the 1970s. 
The number of active neutrino flavors was measured to be three by the
LEP experiments and by SLD.  
However, despite immense experimental efforts, discovery of neutrino
mass and flavor mixing remained beyond reach until deep
underground experiments were carried out to observe neutrinos of
extra-terrestrial origins.  

The study of the fundamental properties of neutrinos is now in a major
discovery phase.  There are good reasons to expect important neutrino
discoveries to continue through the next two decades using new
accelerator based neutrino sources, as will be summarized in this
report.  The rich new physics that can be explored will test the most
basic tenets of particle theory:  the
masses and mixing of fermions and CP violation.  

\section{Evidence for neutrino oscillations \ } 

One crucial breakthrough in neutrino mass studies occurred in the 
measurements of
neutrinos produced in the atmosphere by cosmic ray interactions.  The
decays of pions and muons produced by cosmic rays give a roughly
isotropic flux of neutrinos at energies of a few GeV that has the
composition of about two $\nu_\mu$ to one $\nu_e$.   Instead, the first
experiments to measure neutrinos of atmospheric origin found a
$\nu_\mu/\nu_e$ ratio which was about 60\% of the 
expected value.  This deficit was
soon interpreted as evidence that oscillations occurred between the two
flavors, but to decide which oscillations took place and the mass scale
of the oscillations required larger detectors. The construction of the
SuperKamiokande detector with 22.5 kilotons of 
ultra-pure water fiducial
volume collected large statistics.  A  dependence of the $\nu_\mu/\nu_e$
event ratio on the zenith angle of the neutrino was found, with
down-going neutrino events in agreement with expectation and up-going
events a factor of about two below expectation \cite{kam}.   
This important result,
also confirmed by other experiments \cite{imb}\cite{soudan2}\cite{macro}, 
establishes that muon neutrinos
disappear as the baseline increases, while electron-neutrinos are
observed at the expected rate at all baselines. 
The limits from reactor experiments support the latter inference. The
most economical theoretical interpretation is that $\nu_\mu$'s
oscillate to $\nu_\tau$'s with large mixing
($\sin^2 2\theta \simeq 1$) and mass-squared difference $\delta m^2 \simeq 3\times10^{-3}\rm\,eV^2$.  The
early results from the K2K experiment are more consistent with the
oscillation interpretation than with no oscillations~\cite{k2k} ; the future
MINOS and CNGS
long-baseline experiments have the potential to provide more
precision and may for the first time observe the appearance of 
$\nu_\tau$ or even $\nu_e$ events from oscillations. 

%part 2
Another important breakthrough is that the long-standing solar neutrino problem
is now known to be 
caused by the oscillation of electron neutrinos, at a
mass scale well below that probed by atmospheric neutrino experiments. 
Measured deficits in the solar neutrino flux of 1/2 to 1/3 compared to
Standard Solar Model predictions defied astrophysical explanation. Now,
when the electron neutrino
flux measurement from the SNO experiment is combined with the solar flux
from SuperKamiokande, it is deduced that muon neutrino and tau neutrino
contributions are seen at the 3 sigma level, confirming that neutrino
flavor oscillations have occurred \cite{sno}.  
Global analyses of solar neutrino
data find that the Large Mixing Angle (LMA) solution at 
$\delta m^2 \simeq 5\times10^{-5}\rm\,eV^2$ is
preferred, which is very fortunate
since future long-baseline experiments can probe this mass scale. The
KamLAND experiment is expected to measure both the square of the
$\delta m^2$ and $\sin^2 2\theta$ to 10\% accuracy if the solar solution is 
LMA~\cite{kamland1}\cite{kamland2}.

A complication to a three-neutrino oscillation interpretation is the LSND
data, which give
evidence for $\nu_\mu \to \nu_e$ oscillations at around 
the 1~eV$^2$ scale, with a very small mixing amplitude \cite{lsnd}.  
To have oscillations at three distinct
mass-squared difference
scales, a sterile neutrino with no Standard Model weak interactions must
be invoked. The atmospheric and solar data still allow sterile neutrinos.
The MiniBooNE experiment will test the LSND result in the near future
\cite{miniboone}. 
For most of this
report we concentrate on the three neutrino scenario for brevity, but
even richer
oscillation phenomena may exist.  Sterile neutrinos have also been
invoked in r-process
nucleosynthesis and as warm dark matter.  Their existence is a profound
issue that can only be directly probed in oscillation studies.

The discovery of flavor oscillations has given us a first glimpse of the
physics of
neutrino mass, which has proven to be extremely interesting.  The
approximate bimaximal mixing form of the neutrino mixing matrix was
totally unexpected, and shows us that the physics of the lepton sector
is not a copy of the physics of the quark sector, wherein all mixings
are small.  The immediate challenge before us is to measure the small
mixing angle between the first and third generation neutrino mass
states, since that mixing is vital to neutrino interactions in matter
and to CP violation in the lepton sector.  In
principle, the lepton sector is a better probe of fermion mass physics
than the quark sector, since there are no complications from the strong
interactions.  The tools to develop neutrino mass physics into a
precision science, using intense accelerator neutrino beams and large
detectors, are within reach and the first steps with superbeams can be
made at reasonable cost.  Long baselines are essential to probe both the
atmospheric and solar mass scales, so the experiments necessarily
involve different laboratories and possibly 
even different countries, making this
a truly international scientific enterprise.  More than one facility
will be needed to resolve the fundamental questions before us.

%\label{}
\section{Neutrino Oscillation overview } 
The neutrino flavor eigenstates $\nu_\alpha\ (\alpha = e, \mu, \tau)$
are related to the mass eigenstates $\nu_j\ (j = 1, 2, 3)$ in vacuum by
$\nu_\alpha = \sum U^{*}_{\alpha j} \nu_j \;$,
where $U^*$ is the $3\times3$ mixing matrix. 
It can be parametrized by
\begin{eqnarray}    \addtolength{\arraycolsep}{-.2em}
U^* {=}\!\! \left(\begin{array}{ccc}
c_{13} c_{12}    &  c_{13} s_{12}  & s_{13} e^{i\delta} \\
- c_{23} s_{12} - s_{13} s_{23} c_{12} e^{-i\delta}
& c_{23} c_{12} - s_{13} s_{23} s_{12} e^{-i\delta}
& c_{13} s_{23} \\
    s_{23} s_{12} - s_{13} c_{23} c_{12} e^{-i\delta}
& - s_{23} c_{12} - s_{13} c_{23} s_{12} e^{-i\delta}
& c_{13} c_{23}
\end{array}\right)  %\nonumber\\
  % && \hspace{7em} \times
\hskip-.5em
\left(\begin{array}{ccc}
1& 0& 0\\
0&\  e^{i\phi_2}& 0\\
0& 0& e^{i(\phi_3+\delta)};
\end{array}\right) && \quad
\end{eqnarray}
where $c_{ij}=\cos\theta_{ij}$ and $s_{ij} = \sin\theta_{ij}$. The extra 
diagonal phases are present for Majorana neutrinos but do not affect 
oscillation phenomena. 

The vacuum oscillation probabilities are given by
\begin{eqnarray}
P(\nu_\alpha\to\nu_\beta) &=& -4\Re (U_{\alpha 2} U_{\alpha3}^*
U_{\beta2}^*
U_{\beta3}) \sin^2\Delta_{32} - 4\Re (U_{\alpha 1} U_{\alpha3}^*
U_{\beta1}^*
U_{\beta3}) \sin^2\Delta_{31}\nonumber\\
&& -4\Re (U_{\alpha 1} U_{\alpha2}^* U_{\beta1}^* U_{\beta2})
\sin^2\Delta_{21} \pm 2JS \;,
\end{eqnarray}
where $J = \Im(U_{e2} U_{e3}^* U_{\mu2}^* U_{\mu3})$, 
is the $CP$-violating invariant,
and $S = \sin2\Delta_{21}  + \sin2\Delta_{32} - \sin2\Delta_{31}$, 
is the associated dependence on $L$ and $E_\nu$,
Here, 
$\Delta _{jk} \equiv \delta m_{jk}^2 L/4E_\nu = 1.27 (\delta m_{jk}^2/{\rm
eV^2})
(L/{\rm km}) ({\rm GeV}/E_\nu)$. The plus (minus) sign is used when 
$\alpha$ and $\beta$ are in cyclic (anticyclic) order, where cyclic order
is defined as $e \mu \tau$.
The physical
variable is $L/E_\nu$, where $L$ is the baseline from source to
detector and $E_\nu$ is the neutrino energy.  Only for the LMA
solution is the secondary mass scale sufficiently large that
$CP$-violation can be probed at long
baselines~\cite{bpw,pakvasa,derujula}.

The propagation of
neutrinos through matter is described by the evolution
equation~\cite{matter,kuo}
\begin{equation}
i{d\nu_\alpha\over dx} = \sum_\beta {1\over 2E_\nu} \left( \delta m_{31}^2
U_{\alpha3} U_{\beta3}^* + \delta m_{21}^2 U_{\alpha2} U_{\beta2}^* + A
\delta_{\alpha e} \delta_{\beta e}\right) \;,
\label{eq:evolution}
\end{equation}
where $x = ct$ and $A/2E_\nu$ is the amplitude for coherent forward
charged-current $\nu_e$ scattering on electrons, with
\begin{equation}
A = 2\sqrt 2\, G_F \, Y_e \, \rho\, E_\nu = 1.52\times10^{-4}\,{\rm eV^2}
Y_e \, \rho\,({\rm g/cm^3}) E_\nu\,(\rm GeV) \;.
\label{eq:A}
\end{equation}
Here $Y_e(x)$ is the electron fraction and $\rho(x)$ is the matter density.
In the Earth's crust and mantle, the average density is typically
3--5~gm/cm$^3$ and $Y_e\simeq 0.5$. 
The evolution equations can be solved numerically taking into account the
dependence of the density on depth using the density profile from the
Preliminary Reference Earth Model\cite{prem}.

With three neutrinos there are two independent $\delta m^2$, and 
$|\delta m_{31}|^2 \gg |\delta m_{21}|^2$ is indicated by the atmospheric and 
solar oscillation evidence. A
 recent analysis of the Super-Kamiokande (79 kton-yr) and the MACRO atmospheric
neutrino data finds \mbox{$1.5\times 10^{-3} < \delta m_{31}^2 < 4.5 
\times 10^{-3}$ eV$^2$} and 
$\sin^2 2\theta_{23} > 0.84$ at the 
 95\% confidence level with the best-fit at 
$|\delta m_{31}|^2= 2.7\times 10^{-3}$ eV$^2$ and $\sin^2 2\theta_{23}= 0.96$
~\cite{valle,macro}.
The best fit solution to the latest solar neutrino data, 
$\delta m_{21}^2=4.9 \times 10^{-5}$ eV$^2$ and $\sin^2 2\theta_{12}=0.79$,
lies in the LMA region
with 
\mbox{$2\times 10^{-5} < \delta m_{21}^2 < 2 \times 10^{-4}$ eV$^2$} 
and 
$\sin^2 2\theta_{12} > 0.6$ at the 
 95\% confidence level~\cite{fogli,kam}. Note that the MSW solution 
selects $\delta m_{21}^2>0$. 
The sign of $\delta m^2_{31}$ can be either positive or
negative, corresponding to having the most widely separated mass
eigenstate above or below the other two mass
eigenstates. 
The current generation of nuclear reactor experiments 
(Palo Verde and CHOOZ) 
find null oscillation results
and rule out $\bar{\nu}_{e}\rightarrow \bar{\nu}_x$ oscillations for 
$\delta m_{31}^2 > 10^{-3}$ $\rm{eV}^2$ at maximal mixing and 
\mbox{$\rm{sin}^2 2\theta_{13} > 0.1$} 
for larger $\delta m_{31}^2$ (at the 95\% confidence level)
\cite{paloverde,chooz}. 

 Approximate formulas for the oscillation probabilities in
matter of constant density in
the limit \mbox{$|\delta m^2_{21}| \ll |\delta m^2_{31}|$}, 
have been derived~\cite{cervera,freund,shrock}. 
Expanding in $\alpha \equiv
\delta m^2_{21}/\delta m^2_{31}$, the $\nu_\mu \to
\nu_e$ and $\bar\nu_\mu \to \bar\nu_e$ probabilities for $\delta
m^2_{31} > 0$ are
\begin{eqnarray}
P(\nu_\mu \to \nu_e) = x^2 f^2 + 2 x y f g (\cos\delta\cos\Delta - \sin\delta\sin\Delta)
+ y^2 g^2\,,
\label{eq:P}
\\
P(\bar\nu_\mu \to \bar\nu_e) = x^2 \bar f^2 + 2 x y \bar f g (\cos\delta\cos\Delta
+ \sin\delta\sin\Delta) + y^2 g^2 \,,
\label{eq:Pbar}
\end{eqnarray}
respectively, where $\Delta \equiv \delta m_{31}^2 L/4E_\nu 
= 1.27 \delta m_{31}^2({\rm eV^2}) L({\rm km})/ E_\nu({\rm GeV}) \,,\ \ 
\hat A \equiv A/\delta m_{31}^2 \,$, and
\begin{eqnarray}
x &\equiv& \sin\theta_{23} \sin 2\theta_{13} \,,\ \ 
\label{eq:x}
y \equiv |\alpha| \cos\theta_{23} \sin 2\theta_{12} \,,\ \ 
\label{eq:y}
f, \bar f \equiv \sin((1\mp|\hat A|)\Delta)/(1\mp|\hat A|) \,,\ \  
\label{eq:f}
g \equiv \sin(|\hat A\Delta|)/|\hat A| \,.
\label{eq:g}
\end{eqnarray}
Note that the existence of matter effects ($A\neq 0$) makes it possible
to discriminate between $\Delta >0$ and $\Delta < 0$, and thereby determine
if there are two heavy mass eigenstates or just one (assuming the mass 
differences themselves are on the order of the highest mass eigenstate).  
The corresponding probability for a $T$-reversed channel is found by
changing the sign of the $\sin\delta$ term.
The formulas are valid at $E_\nu > 0.5$ GeV and $L <4000$ km 
for all values of
$\delta m^2_{21}$ currently favored by solar neutrino experiments. The
corresponding expansion in $\alpha$ and $\theta_{13}$ in vacuum can
be found by the substitutions
$\sin((\hat A - 1)\Delta)/(\hat A - 1) \to \sin\Delta$ and
$\sin(\hat A \Delta)/\hat A \to \sin\Delta$. Approximate analytic expressions
for the probabilities for low energy beams have been derived in 
Ref.~\cite{arafune}.

Neutrino oscillations can probe violations of
 the discrete symmetries $CP$, $T$ and $CPT$. $CPT$ invariance is a basic 
property of local quantum field theory and no deviations from it have been
found to date, but $CPT$ non-conservation may occur in string theory. 
If $P(\nu_\alpha \to \nu_\beta) \neq {P}(\bar\nu_\beta \to \bar\nu_\alpha)$ or
$P(\nu_\alpha \to \nu_\alpha) \neq {P}(\bar\nu_\alpha \to \bar\nu_\alpha)$, 
then $CPT$ is violated. If 
$P(\nu_\alpha \to \nu_\beta) \neq {P}(\bar\nu_\alpha \to \bar\nu_\beta)$, 
$CP$ is not conserved. If 
 $P(\nu_\alpha \to \nu_\beta) \neq P(\nu_\beta \to \nu_\alpha)$ then $T$
invariance is violated.  When neutrinos propagate through matter, 
fake $CP$ and $CPT$ violation effects may be observed 
even if the mass matrix is
 $CP$ conserving. However, matter-induced $T$-violating effects are negligible
and are completely absent if the matter density-profile is symmetric 
with respect to the locations of the source and detector.

\section{The Need for New Facilities and New Detectors}

The first imperative for the currently planned experiments is that 
they achieve their primary goals for sensitivity to 
$\sin^2 2\theta_{23}$ and $\delta m_{23}^2$.  The next imperative is the 
detection of $\nu_\mu$ to $\nu_e$ at a baseline corresponding 
to the atmospheric mass splitting, 
and the determination of the angle $\theta_{13}$, for which there
presently exists only an upper limit. The MINOS and CNGS 
experiments will have sensitivity down to $\sin^22\theta_{13} = 0.02$,
and the JHF proposal is slightly more sensitive (and will be discussed
more in detail later), 
but the appearance amplitude could be much smaller.  In fact, some 
theoretical models suggest that $\sin^22\theta_{13}$ could be at the
$10^{-3}$ to $10^{-4}$ level \cite{albrightgeer}.  
The value of $\theta_{13}$ is crucial
to differentiate theoretical models of neutrino mass generation.  Only
after $\theta_{13}$ is measured can CP studies and the determination of
the sign of $\delta m^2_{13}$ be achieved.  

The logical path to increased sensitivity at a reasonable cost is to
upgrade conventional beams to higher intensity and concurrently
construct a large underground detector at a suitable
distance.  The energy of the neutrino superbeam and the
baseline to the detector must be selected to optimize the physics
reach~\cite{super}.  For the $\nu_\mu$ to $\nu_e$ oscillation to be
nearly maximal, the average neutrino energy at a given baseline $L$
should be chosen such that $\Delta_{31} = (2n+1)\pi/2$.  This choice also
makes the $\nu_\mu$ to $\nu_\tau$ oscillation maximal, which aids in
$\tau$-appearance studies with a neutrino energy well above the
$\tau$-threshold energy of 3.56~GeV.  Additionally, for CP studies,
the $\delta$ dependence here is pure $\sin\delta$, with no 
$\cos \delta$ contribution, which eliminates a
$\theta_{13}$--$\delta$ ambiguity.  The above statements remain valid
even in the presence of matter.  A narrow band neutrino beam is
advantageous in order to have all neutrinos near the same $L/E_\nu$
value and to eliminate backgrounds, which are significant in 
conventional beams.  

To learn the most physics for the given 
investment of a large neutrino detector, it is important that the
detector be chosen and housed such that it can also be used to study other 
phenomena.  Important areas of study which could also benefit from 
a new large underground detector include atmospheric and solar
neutrino studies, and searches for both proton decay and neutrinos from 
supernovae.  
The lower the detector energy threshold, and the deeper underground
the detector, the more physics it can access.  Since we do not know where 
the next hint of physics beyond the Standard Model will lie, it is 
important to keep as many avenues open as possible.  

The following sections of this report describe a set of conventional 
neutrino beamlines and detectors that could be used to take the 
next step in neutrino oscillation physics.  Where relevant, other
non-accelerator physics possibilities will be mentioned.  The 
physics capabilities for these conventional beamlines will be 
described, followed by a discussion of the capabilities
of a neutrino factory.  In fact, what new measurements the neutrino 
factory can provide depend critically on what the next few years of neutrino 
experimentation will tell us:  What are the parameters of the solar
neutrino oscillation?  Is there a sterile neutrino sector?  Is the 
small mixing angle between the third and first generation more than a 
per cent or so?  We conclude this section with a discussion of 
the different possible answers to these questions, and what we would learn 
in each case from a neutrino factory.  

\subsection{Current Superbeam Proposals}

There are currently three superbeam proposals which involve new 
neutrino beamlines directed towards large underground water \u{C}erenkov
detectors.  Because of the wealth of experience now with water \u{C}erenkov
detectors at extremely low neutrino energies (at the sub-GeV level), 
and the fact that a 50 kton detector already exists, the reaches of these
proposals have been evaluated with known detector effects and beam-related
backgrounds.  Table 
\ref{tab:sb1} shows a summary of the physics reach of these proposals.  
In summary, the JHF-based proposals involve building a beamline which 
can focus different momentum pions to make a narrow band beam to run
at the oscillation peak, as is suggested above \cite{jhf}.  
The CERN to UNO proposal
starts with a much lower energy proton beam and has inherently lower
backgrounds, so they can reach comparable sensitivities with a wide band
low energy neutrino beam \cite{cernuno}.  

\begin{table} 
\caption{Physics Sensitivity for Current Superbeam Proposals} 
\label{tab:sb1} 
\begin{tabular}{llllll}
\hline\hline
Name & Years of Running & kton & 
$\sin^2 2\theta_{13}$ & CP Phase $\delta$ & $\nu$ Energy \\ 
  &   &   &  sensitivity (3$\sigma$) & sensitivity (3$\sigma$) & (GeV) 
\\ \hline
JHF to SuperK & 5 years $\nu$ & 50 & 0.016 & none & 0.7 \\ \hline
SJHF to HyperK & 2 years $\nu$, 6 years $\bar\nu$ & 1000& 0.0025 & $>15^\circ$  & 0.7 \\ \hline 
CERN to UNO & 2 years $\nu$, 10 years $\bar\nu$ & 400 & 0.0025 & $>40^\circ$ 
& 0.3 \\
\hline
\end{tabular} 
\end{table} 

The JHF project involves building a new
neutrino beamline at the Japanese Hadron Facility and aiming it toward
the SuperKamiokande detector, at a distance of 295\,km.  
This project has a few different 
models for narrow-band beams, each starting with the new 50\,GeV proton
source.  The JHF facility itself has begun construction and is expected to 
finish in 2007.  The neutrino beamline could conceivably begin operations
by 2008, and is being designed to receive protons from a 0.77\,MW proton 
source.  Assuming that $\delta m_{32}^2$ is above $1\times 10^{-3}eV^2$, 
this facility could see $\nu_\mu \to \nu_e$ at the three $\sigma$ level 
if $\sin^2 \theta_{13}$ is larger than 0.016, and after five 
years of data-taking.

The JHF neutrino program also 
has an upgraded proposal, whereby both the beamline and 
the detector is augmented.  The proton source itself would  
be upgraded to 4\,MW, and the beamline elements would have to be
fortified accordingly.  The new detector proposed, HyperKamiokande, 
is another water \u{C}erenkov device, but with 20 times the fiducial mass 
of the SuperKamiokande detector, located under the same mountain as 
the SuperKamiokande detector.  With this increase of 100 in exposure 
(in terms of kton-years) they expect to be able to see $\nu_\mu \to \nu_e$
at 3$\sigma$ if $\sin^22\theta_{13}$ is larger than 0.0025.  If 
$\sin^22\theta_{13}$ is 0.01, and $\delta m_{21}^2$ is $1\times 10^{-4}eV^2$, 
then CP violation could be observed if the CP phase $\delta$ is above 15
degrees (assuming the expected mass hierarchy for matter effects), 
assuming a two year neutrino run, and a six year antineutrino run.  

Finally, there is a proposal which makes use of the Superconducting Proton 
Linac at CERN, which will be a 2.2\,GeV proton source, operating potentially
at 4MW.  The 2.2\,GeV proton beam will be used to make a 300\,MeV neutrino 
beam, with very low (and variable) intrinsic $\nu_e$ background.  By 
aiming this beam at a cavern in the Frejus tunnel, located 150\,km away, 
they could measure $\nu_\mu\to \nu_e$ if $\sin^22\theta_{13}$ is larger 
than 0.0025, similar to the SJHF to HyperK proposal.  CP violation could
be observed if the phase $\delta$ is larger than 40 degrees (for the 
same parameters mentioned above)  
assuming a two year neutrino run, and a ten year antineutrino run.  

\subsection{Choosing a Neutrino Energy} 

     The choice to use low energy neutrino beams in the experiments 
in table \ref{tab:sb1} was motivated by several factors:  
for the CERN to UNO proposal, the intense proton source will be at 
2.2GeV, requiring a very low energy beam, which has in turn a very 
low electron neutrino background.  Also for these experiments, the performance
of Water \u{C}erenkov at these energies is very well understood.  In the 
following section a case will be made for more than one superbeam, 
and in particular one at significantly higher energies than these 
two proposals.  It is an important exercise, however, to understand
how different energy beamlines compare.  In the following section, it is 
assumed the flux achievable at one baseline is constant with energy as an 
exercise. 

\begin{figure}
\includegraphics[width=\linewidth]{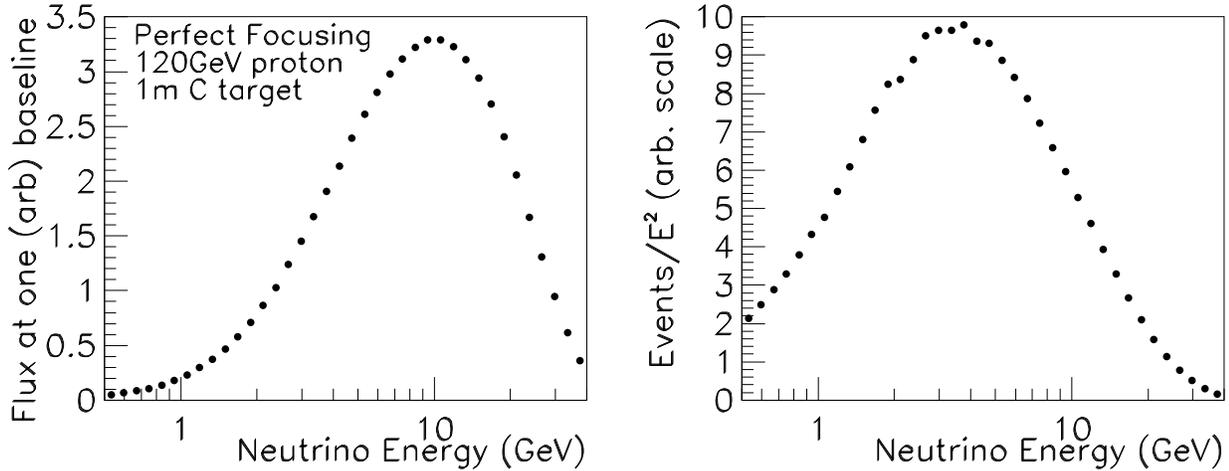} 
\caption{Fluxes (arbitrary normalization) for a series of  
perfectly focused ``monochromatic'' neutrino beams, 
coming from 120GeV protons striking a 1m long graphite target.  The
left plot shows the flux at one baseline for several different focused
energies, and the right plot shows the unoscillated 
event rate comparison (arbitrary scale) scaled by $1/L^2$, at the 
oscillation peak, or ($1/E_\nu^2$).} 
\label{fig:pbeam}
\end{figure}

     To produce an on-axis narrow band neutrino beam one focuses a given 
momentum bite of pions.  However, for different momentum bites, one
would want to direct the beam to different baselines.  Figure \ref{fig:pbeam}
(left) shows the event distribution at 730\,km coming from 
for a perfectly focused ``monochromatic'' pion beams, 
which were produced by 120\,GeV protons striking
a 1\,m graphite target \cite{pbeam}.  
To understand how the unoscillated fluxes would
scale for perfectly focused narrow band beams, the 
distribution is shown in fractional energy bands (i.e. assume
a perfectly focused pion beam with a constant fractional momentum bite 
at each energy).  The most useful
way to understand the beamline 
capability as a function of energy is to divide the 
event rate for that momentum bite by the square of the energy, since if one 
is operating at a given multiple of the peak 
the baseline ($L$) scales like the energy, and of course
the event rate for these baselines scales like $1/L^2$.  The result
of that operation is shown in figure  \ref{fig:pbeam} on the right, for 
the same target and proton beam.  As the pions one is trying
to focus get higher in energy 
they are boosted more forward, so the 
far detector sees even more events than the $1/L^2$ scaling takes away. 
However, once the $\nu$ energies are above about 5\,GeV the scaled rates
drop, because much fewer high energy mesons are produced by  
120GeV protons.  
Also, the finite length of the decay tunnel will also affect
the highest energies--for the plot shown here the entire beamline 
is assumed to be 
725m long.  Clearly one also needs to understand how efficient the 
focusing can be as a function of pion energy, as well as the possibilities
for off-axis neutrino beams, which can also provide 
narrow neutrino energy spectra.  

\subsection{The case for more than one Superbeam} 

The measurement of the transition of $\nu_\mu \to \nu_e$ is 
extremely important and helps link the solar and atmospheric 
anomalies together.  Although any one superbeam measurement would signal 
a breakthrough in our understanding of the mixing matrix, 
the nature of superbeams is such that the signal will not be
background-free, and the interpretation of the result will depend
significantly on the ability of any experiment to predict its background.  
Furthermore, while the experiments outlined above have impressive 
sensitivity to the small mixing angle itself, they have no sensitivity
to the sign of the largest mass splitting.  In order to 
get the most information out of the $\nu_\mu \to \nu_e$ transition, 
it is important that there be an experiment which is sensitive to 
matter effects.  Superbeams in principle may be able to measure matter 
effects if $\theta_{13}$ is large enough, but not the programs 
outlined above.  
Finally, the SJHF-HyperKamiokande proposal shows that matter effects 
are the same size as a change in the phase $\delta$ of about 8 degrees, 
in the LMA scenario.  If that experiment does indeed see evidence for 
CP violation before the sign of matter effects is known, then there 
will be an additional uncertainty in whether or not CP violation in 
the lepton sector does actually occur.  

Extensive studies have already been made of superbeams from BNL and
Fermilab upgrades \cite{fnalsuperbeams}.  
For the discussion below we focus on physics
results~\cite{bmw} that could be achieved with a 1.6~MW proton driver at
Fermilab to obtain a factor of four intensity increase over the NuMI
medium energy beam. Similar fluxes could be obtained with a BNL
superbeam, so our conclusions should apply generally to either
facility.  To represent the anticipated flux loss in making a narrow
band neutrino beam, we divide the flux estimate by a factor of five.
Both liquid argon and megaton water \u{C}erenkov detectors are under
active consideration, and similar sensitivities to oscillation physics
can be achieved with either detector. Our examples of the physics
reach below are based on a liquid argon detector with an effective
70~kt-yr of data accumulation for detecting $\nu_e$'s (e.g., a 70\,kt
liquid argon detector with 50\% efficiency with 2 years of
running). For $\nu_\tau$ detection a 3.3\,kt-yr exposure will be
assumed (e.g. a 5\,kt detector with 33\% efficiency and 2 years of data
taking).  Antineutrino event rates are three to six times lower than
neutrino rates, and correspondingly longer running times are required
to accumulate comparable numbers of events.  
We assume a $\nu_e$ fractional background ($f_B$) (which includes
both detector and beam backgrounds) of 0.4\% of the unoscillated CC
signal, and a fractional uncertainty on the background
of $\sigma_{f_B}/f_B$ of 10\%.  For concrete illustrations, we assume
$\delta m_{31}^2 = 3.5\times10^{-3}\rm\,eV^2$, $\theta_{23}=\pi/4$,
$\delta m_{21}^2 = 5\times10^{-5}\rm\,eV^2$, $\theta_{12} = 0.55$,
recognizing that the physics reach is somewhat dependent on the
oscillation parameters.  Matter effects are taken into account.  A
range of baselines from 350~km to 2900~km is considered, with the
corresponding optimal energies from 1~GeV to 8.2~GeV, respectively.

\begin{figure}
\includegraphics[width=.5\linewidth]{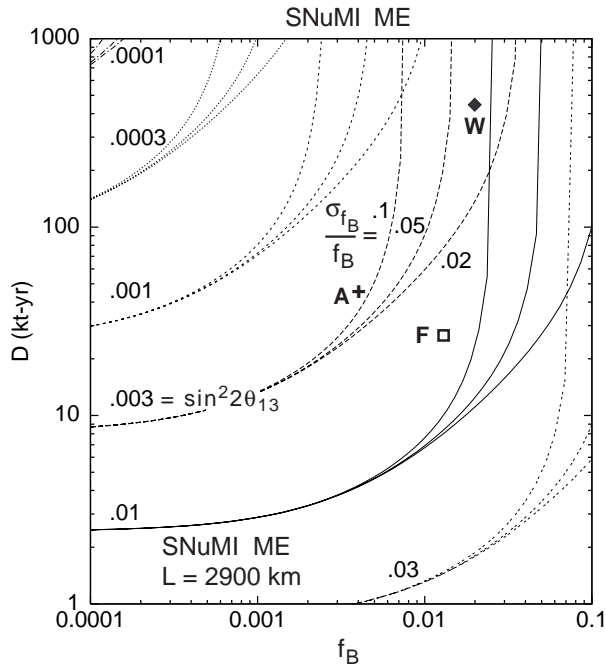} 
\caption{Contours of constant
$\sin^2 2\theta_{13}$ reach that correspond to a $\nu_\mu \to \nu_e$
signal that is $3\sigma$ above background for an upgraded NuMI medium
energy beam with detector at $L= 2900$~km. The contours are shown in the
$(D,f_B)$--plane, where $D$ is the data--sample size. Curves are shown
for systematic uncertainties $\sigma_{f_B}/f_B =$~0.1, 0.05, and 0.02.
Three typical detector scenarios are shown: A (liquid Argon), F
(steel--based), and W (water cerenkov) (from Reference 17). } 
\label{fig:newbk}
\end{figure}

%part 4
In this superbeam scenario, the neutrino event rate in two years
running at a 350~km baseline (e.g., BNL to Cornell) would be 15 times
that expected for 5 years running in the JHF to SuperKamiokande
experiment at a 295~km baseline.  Hence, superbeams offer a dramatic
improvement in the $\sin^22\theta_{13}$ reach.  The $\nu_\mu$ to
$\nu_e$ appearance sensitivity at $3\sigma$ for baselines from 350~km
to 2900~km is, respectively, 0.002 to 0.003, an order of magnitude
below the reach of the upcoming long baseline experiments with
conventional beams.  Figure \ref{fig:newbk} shows the typical
sensitivity of the $\sin^22\theta_{13}$ reach to $f_B$ and
$\sigma_{f_B}/f_B$.  The sensitivity of a
liquid argon or steel--based detector is best improved by increasing the
size, while for a water \u{C}erenkov detector it is best improved by
lowering the backgrounds and/or systematic uncertainty on the
background. The CP phase $\delta$ can be measured
down to 40 degrees at $3\sigma$ for $\sin^22\theta_{13} =
0.01$. However, at the short 350~km baseline, matter effects are
small, so there is very little sensitivity to the sign of $\delta m^2_{13}$.

At baselines above 1200~km, matter significantly affects appearance
rates and the neutrino mass-hierarchy can be resolved.  For example,
at 1290~km (e.g., Fermilab to Homestake) the sign of $\delta m^2_{13}$
can be determined at a three $\sigma$ reach of 0.01 on
$\sin^22\theta_{13}$, which improves by a factor of two at 2900\,km.
Moreover, due to the higher optimal energies for long baselines, $\nu_\tau$
appearance studies become feasible at distances of 1290\,km and above.
For example, $14$ $\nu_\tau$ events would be observed at 1290\,km ($77\nu_\tau$
events at 1770\,km, FNAL-Carlsbad or BNL-Soudan).  However, there is
no sensitivity to the CP phase at these distances.

A novel idea for a superbeam which 
resembled the front end of a neutrino factory was proposed at this workshop.  
With all the proposals
outlined above, the assumption is that the secondary beam focusing will 
be done primarily with horns, and so the beam will be either neutrinos
or antineutrinos.  For the target and focusing of a neutrino factory, 
there will be solenoidal focusing and therefore a mixed $(\nu + \bar\nu$) 
beam.  If the 
far detector could measure the outgoing charge of the electron (and muon) then
it could search for $\nu_\mu \to \nu_e$ and $\bar\nu_\mu \to \bar\nu_e$
simultaneously.  Also, a solenoidal focusing system would have a much lower
failure rate than a horn system 
at a high proton power source, since there would be 
no material in the secondary beam.  
Recent studies have shown that a liquid argon calorimeter might be able to 
measure the charge of outgoing electrons up to a GeV \cite{laxsec}, 
so for a low energy solenoid-focused neutrino superbeam, this could 
be an attractive option \cite{kahn}.  A superbeam detector which could
identify lepton charge could then also be used as the far detector 
for a neutrino factory beam.  

Furthermore, longer baselines (for example, those where 
$L_n=(2n+1)\frac{2\pi E_\nu}{\delta m_{31}^2}$)
with superbeams may be useful for studies of CP 
violation, due to the fact that the asymmetry increases linearly 
with baseline length in the small $\delta m_{21}L/E$ approximation
\cite{xtralbl}.  If the asymmetries one is trying to measure are 
larger, then fixed fractional uncertainties on background fluxes, 
detector acceptances, etc, become less important.  The challenge 
in that case is to provide an intense enough neutrino flux to produce
enough events in the far detector, and to make sure that matter 
effects do not enter in so much as to obscure the asymmetry.  

There is thus a complementarity of superbeam experiments at short and
long baselines.  Experiments at the first oscillation maximum can search for 
the oscillation itself and are most sensitive to seeing 
$\theta_{13}$ if it is very small.
Experiments at higher multiples of the oscillation maximum 
may be able to test for CP
violation in the lepton sector.  Experiments at higher energies and 
long baselines can determine the neutrino mass hierarchy and 
measure $\nu_\tau$ appearance.  We conclude that the full physics of the
neutrino sector can only be explored with superbeams at short and long
baselines.

\subsection{Detector options for superbeams}

The first two requirements for a neutrino detector for a superbeam is that 
it can identify both $\nu_\mu$ and $\nu_e$ charged current interactions, 
and measure the total neutrino energy of the neutrino interaction.  
At lower neutrino energies the neutrino cross section is primarily
quasielastic, so the detector simply must identify an outgoing muon 
or electron.  However, for neutrino energies above a GeV there is 
substantial hadronic activity accompanying most neutrino scattering
events, and the difference between the detectors discussed below lies
in their abilities to identify that accompanying hadronic activity.  

It is important, however, that these large superbeam detectors also
contribute to other important areas of physics, since they will no 
doubt be costly devices.  By housing these detectors deep underground,
they could also make advances in the search for proton decay, supernovae
detection, and solar and atmospheric neutrino studies.  Since we do not
know where the next physics beyond the Standard Model will surface, it 
is important that we not neglect these other areas.  

\subsubsection{Water \u{C}erenkov} 

Water \u{C}erenkov detectors have been the most studied devices 
for superbeams, 
since they have provided some of the most convincing signals for neutrino 
oscillation.  Also, SuperKamiokande is the most massive neutrino detector
constructed to date that would work in a superbeam.  Although they have 
been proven to work extremely well at neutrino energies below 1\,GeV, it 
remains to be seen how well the detector concept would work for higher
energy neutrino beams.  The largest uncertainty is how well they could 
reject backgrounds from higher energy neutrino neutral current 
interactions which contain energetic $\pi^0$'s which decay asymmetrically,
producing an electromagnetic-like ring in the detector.  There is much 
work continuing on different techniques for \u{C}erenkov light collection, 
as well as different techniques for focusing the light itself to improve
the signal.  

A novel idea proposed at this workshop 
is similar to a segmented iron calorimeter but
uses water as the sensitive material. The \u{C}erenkov light is
reflected inside a long thin water tank oriented transversely to the 
incoming neutrino direction (a tank could be $1m\times 1m \times 10m$, and a 
calorimeter would consist of several hundred of these tanks).
and is collected at the end of the tank by small photomultipliers.
Such a detector might be built at a moderate cost compared to a steel-based
detector. A water tank prototype has been built at IHEP, 
Beijing, and tests with cosmic rays are underway.
Preliminary Monte Carlo study shows that its performance is similar to
or better than those of steel-based calorimeters or \u{C}erenkov ring imaging
detectors, particularly at energies higher than a few GeV.
Unfortunately the limited number of photoelectrons for this configuration
prevents it from
being used for low energy physics, such as solar neutrino studies, 
in contrast to the single-volume water \u{C}erenkov devices. 

\subsubsection{Liquid Argon Calorimeter} 

A liquid argon TPC, such as the one being built by the ICARUS collaboration, 
would be an extremely powerful device to use for a neutrino superbeam.  
The detector is basically an electronic bubble chamber, and would be 
able to detect individual tracks in the hadronic showers.  Studies based
on GEANT simulations of the detector show that the neutral current 
background could be suppressed by three orders of magnitude, simply by
looking at the energy loss in the first few radiation lengths of the 
electron candidate in the neutrino event.  Because of the superior vertexing,
most shower-related backgrounds vanish.  There is currently half of a 600\,ton
module of
instrumented liquid argon taking data on cosmic rays, and the data there
look promising \cite{pavia}.  Although the readout cost and cryogenics
prohibit detectors on the 500\,kton scale, as have been proposed 
with water cerenkov detectors, 
the improved signal reconstruction and background rejection 
may provide a high enough signal 
efficiency to make the technology competitive.  

One important unknown about the liquid argon technique is how large 
a single volume could used.  The ICARUS proposal now has 600\,ton 
modules, but to avoid being prohibitively expensive, a much larger
module size must be achieved.  If one could instrument a volume the size
of the SuperKamiokande detector with liquid argon, then it would be 
70\,ktons.  If the signals could then be made to drift across 5\,m then 
a detector this massive would not be prohibitively expensive \cite{lannd1},
\cite{lannd}.  
Discussions with mining engineers have begun and have been encouraging
\cite{brierley}.  
Furthermore, a small volume of this detector should be placed in a neutrino
beam shortly both to measure neutrino cross sections \cite{laxsec}, 
and to understand how closely the actual performance mirrors the 
Monte Carlo prediction.  Finally, if a magnetic field could be introduced
in this detector, then it could be used for the solenoid-focused proposal
discussed earlier, but would ultimately make an ideal detector for a 
neutrino factory beam.  Its low detector threshold and particle 
identification make it particularly attractive for proton decay and 
supernovae searches.  

\subsubsection{Steel-Based Detector} 

A steel-based detector is the most coarse-grained of the detector
options being considered for neutrino superbeams.  Although they 
have typically been used for higher energy neutrino beams, with enough
transverse and longitudinal segmentation they too can provide 
discrimination between $\nu_\mu$ and $\nu_e$ charged current events.  
They typically have neutral current rejection on the order of 
a few per cent; additional kinematic cuts must be used to reduce
that background to the few times $10^{-3}$ level \cite{steelnue}.  
In order to make them particularly interesting for atmospheric 
neutrino studies, they would be magnetized, allowing atmospheric
neutrino studies to be performed on $\nu_\mu$ and $\bar\nu_\mu$ 
separately.  Unfortunately, however, steel-based detectors have a 
detector threshold which would prevent them from being used for solar
neutrino studies or proton decay searches.

\subsection{Physics Reach of a Neutrino Factory} 

It is extremely likely that a superbeam facility would not completely 
determine
the neutrino mixing matrix or measure the CP violating phase.  In that
circumstance a neutrino factory, a natural progression from a
superbeam, will be needed to provide intense beams.  This ultimate
neutrino source will enable precision measurements of the crucial
remaining parameters that could test Grand Unified and other theories
of neutrino mass.  The physics of flavor is one of the major
unanswered problems in particle physics, and complete knowledge of the
flavor changing neutrino processes is essential in developing the
theory of flavor violations.

The neutrino factory concept is to create a millimole per year muon
source, rapidly accelerate the muons to the desired energy, and then
inject them into a storage ring with a long straight section that is
directed towards a far detector.  The decays of the muons in the
ring give $\nu_e$ and $\bar \nu_\mu$ beams for stored positive muons
and $\bar \nu_{e}$ and $\nu_\mu$ beams for stored negative muons.
Thus, for the first time, intense electron neutrino beams would be available.
The resulting neutrino beams will have an energy spectrum that
is well understood from the kinematics of muon decay.  The $\nu_e \to
\nu_\mu$ appearance channels, which lead to wrong-sign muons, yield
relatively background-free signals, above a muon production threshold
of about 4 GeV, which in turn mandates a minimum stored muon energy of
20 GeV.  For such energies, long baselines ($>1800$ km) are optimal
for oscillation studies. An entry level neutrino factory may have 20
GeV stored muons with 10$^{19}$ muon decays in the beam-forming
straight section.  A high performance factory would be a 50 GeV ring
delivering 10$^{20}$ muons per year, yielding 10$^{22}$ kt-decays
after a few years of running. A 50 kt iron scintillator target is
nominally considered for the detector.

\begin{figure}[h]
\includegraphics[width=10cm]{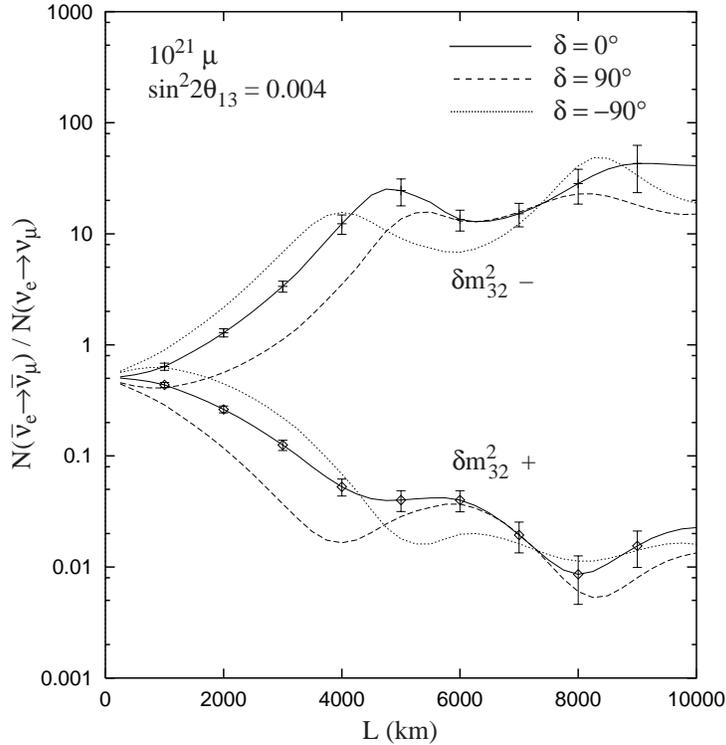} 
\caption[]{The ratio of event rates at a 20 GeV neutrino factory for 
$\delta=0, \pm \pi/2$. 
The upper group of curves is for $\delta m^2_{32} <0$,
the lower group is for $\delta m^2_{32} >0$ and the statistical errors 
correspond to $10^{21}$ muon decays of each sign and a 50 kt detector. The
oscillation parameters correspond to the LMA solution with 
$|\delta m^2_{32}|=3.5 \times 10^{-3}$ eV$^2$ and $\sin^2 2\theta_{13}=0.004$. 
See Ref.~\cite{bgrw}.\label{fig1}}
\end{figure}

What new physics can be explored at a neutrino factory? First, a
neutrino factory could measure $\sin^22\theta_{13}$ down to 10$^{-4}$.
If this mixing angle is indeed below 10$^{-3}$, a factory is the only
way to measure it.  Second, with a detector located at a long
baseline, both the sign of $\delta m^2_{13}$ and the CP phase can be
determined.  Interestingly, the intrinsic CP violating effects are
absent at 7300 km (e.g., Fermilab to Gran Sasso) and maximal CP
violation occurs at a baseline of about 
2900\,km~\cite{barger,campanelli,lindner}, as shown in Fig.~\ref{fig1}.  The
sign of $\delta m^2_{13}$ can be determined via matter effects down to
$\sin^22\theta_{13}$ of 10$^{-4}$.  If $\sin^22\theta_{13}$ is 0.01 
and $\delta m_{12}^2$ is $1\times 10^{-4}$, as was assumed for the 
sensitivities quoted in table \ref{tab:sb1}, then a neutrino factory 
can in three years $\mu^+$ running and six years $\mu^-$ running, 
see a three $\sigma$ CP violation effect down to $\delta=12^\circ$
with a 50kton detector.  However, if $\sin^22\theta_{13}$ is as low
as $2\times 10^{-4}$, then CP violation can still be detected 
at three $\sigma$ if $\delta$ is above $40^\circ$.  

\begin{figure}[h]
\begin{minipage}[b]{8cm}
\includegraphics[width=8cm]{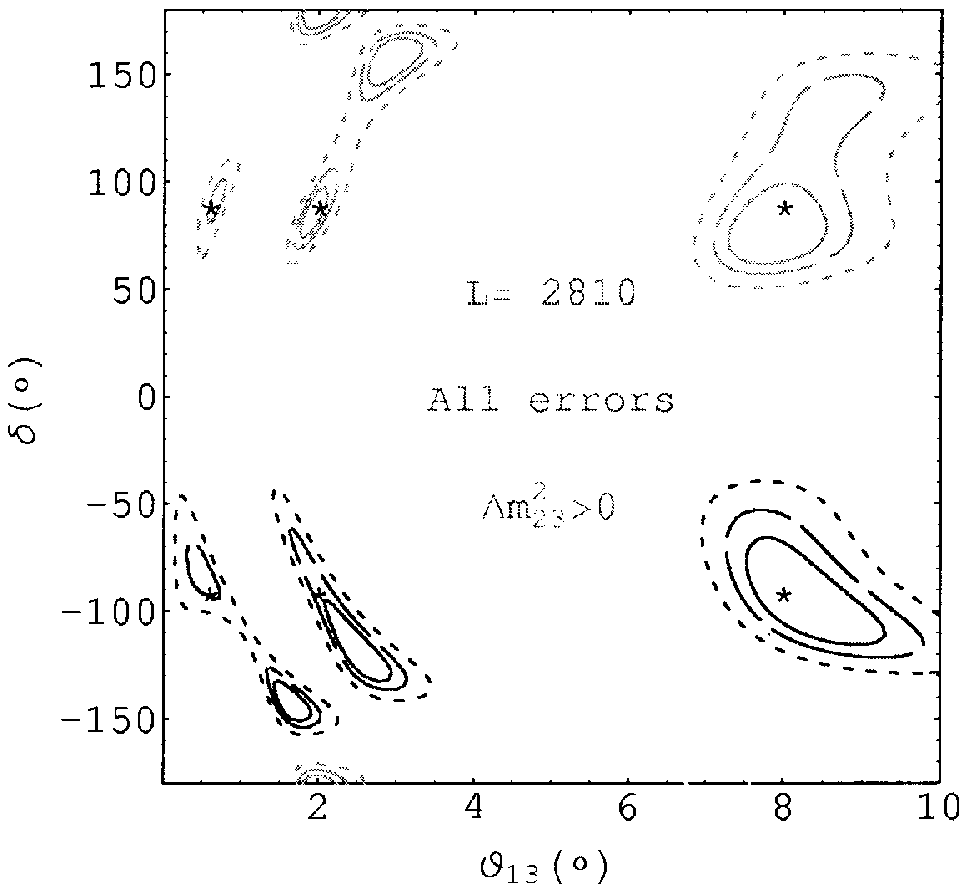} 
\caption[]{Fits of $\delta$ and $\theta_{13}$ for various central values of 
$\delta$ and $\theta_{13}$ in simulations of neutrino factory capabilities 
at $L=2810$ km. 
See Ref.~\cite{castell}.\label{fig2}}
\end{minipage}
%\end{figure}
\hspace{.5cm}
%\begin{figure}[h]
\begin{minipage}[b]{8cm}
\includegraphics[width=8cm]{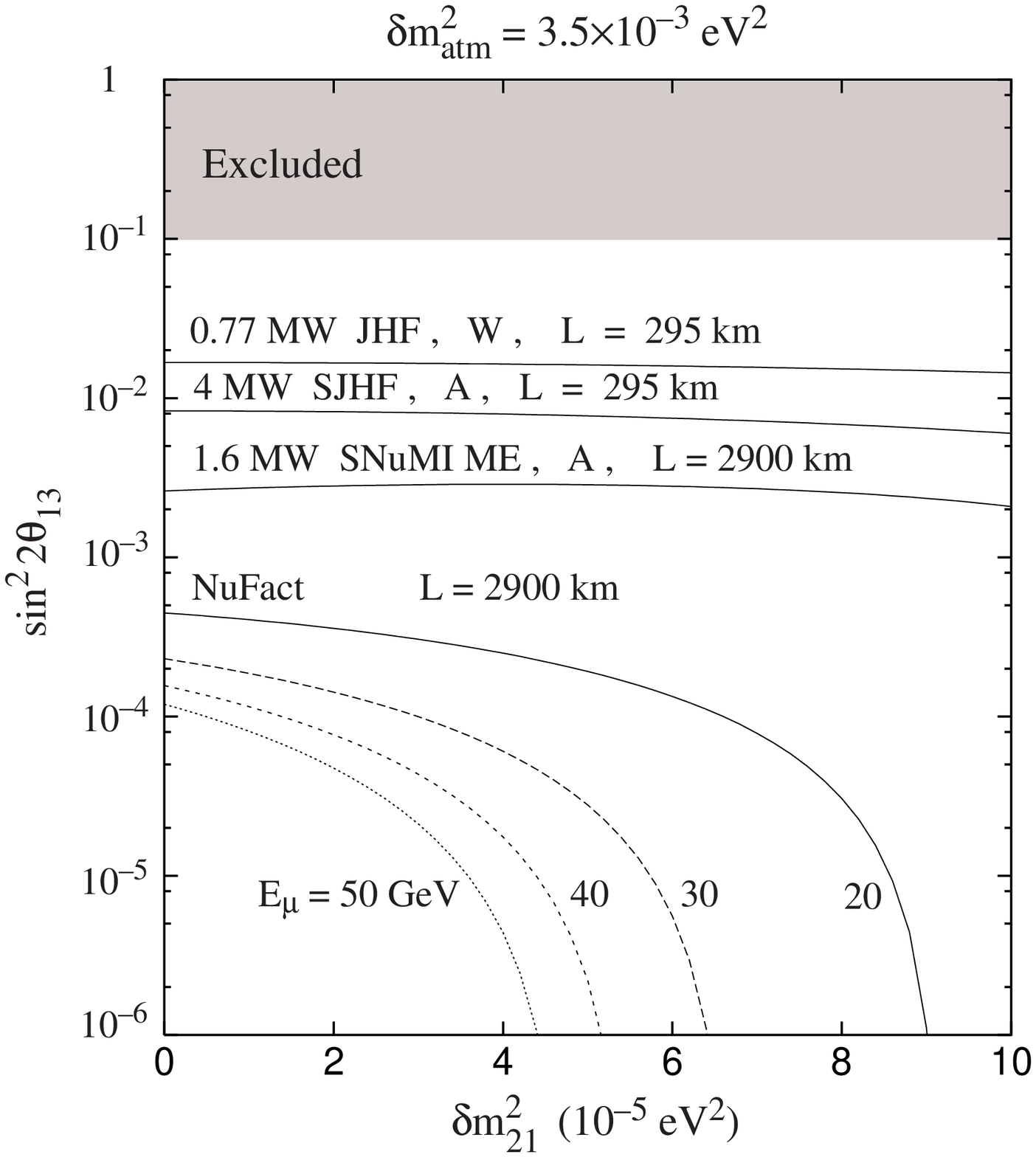} 
\caption[]{A comparison of the physics reach of superbeams and 
neutrino factories.
Adapted from the study in Ref.~\cite{super}.\label{fig3}}
\end{minipage}
\end{figure}

Simulations~\cite{cervera,lindner,yasuda} have been made that
demonstrate determinations of all the oscillation parameters,
including the CP phase, to impressive accuracies.  Figure~\ref{fig2}
shows representative results from one such study.  Figure~\ref{fig3}
compares the physics reach of superbeams and neutrino factories in the
parameters $\sin^2 2\theta_{13}$ and $\delta m^2_{21}$.
The strength of the neutrino factory lies in the precision 
achievable if $\sin^2 2\theta_{13}$ is large, and in the reach 
of the mixing angle itself if $\sin^2 2\theta_{13}$ is small.  

\subsection{Scenarios Leading to Neutrino Factory Measurements} 

It is important to understand what a neutrino factory has to offer 
depending on what scenarios turn out to be true in neutrino oscillation 
physics.  
Keep in mind that at the time of this document we only know that there
are three indications of oscillations 
with non-overlapping mass splittings, and that for at least one of those
mass splittings, the corresponding mixing angle appears to be large.  

The physics that a neutrino factory can provide depends on a few key 
factors, all of which will be determined within the next 5 years or so: 
\begin{enumerate} 
\item whether or not the LSND signature is due to oscillations;
\item if the solar neutrino anomaly is described by the LMA solution; and
\item if the next generation of neutrino experiments sees $\nu_\mu$ to 
$\nu_e$ (in other words, if $\sin^22\theta_{13}$ is up to a factor of 3 
below the current limit from CHOOZ)
\end{enumerate} 

If the LSND signature is in fact due to oscillations, then a neutrino 
factory is basically the only way we can access the sterile neutrino 
sector.  In this scenario it becomes extremely important to try to measure
every transition as accurately as possible, to try to understand the 
structure of the sterile neutrino sector:  is there more than one sterile
neutrino?  High energy neutrino beams will be needed at short baselines 
to study $\nu_\mu$ and $\nu_e$ transitions to $\nu_\tau$ at the LSND 
mass difference, as well as the antineutrino transitions.  It becomes
possible to observe CP violation 
in $\nu_\mu$ to $\nu_e$ at atmospheric mass difference baselines.  
The requirements for the short baseline experiments
are much more modest in this scenario, and one can already start defining
the sterile neutrino sector with an order of magnitude fewer muon decays
per year than what has already been described in study II \cite{study2}.  
MiniBooNE will have adequate sensitivity over the entire LSND signal region
to determine conclusively whether or not we are in this 
scenario \cite{miniboone}.  

If the LSND signature is not due to oscillations, then the next question
that defines the scope of a neutrino factory is whether or not the solar
neutrino anomaly is described by the large mixing angle solution, as is
currently the most favored region.  This will be determined by the KAMLAND
experiment, again in just a few years from now.  

If KAMLAND does see neutrino oscillations, then whether or not CP violation 
is accessible depends on the value of the smallest mixing angle, 
$sin^2 2\theta_{13}$.
If the next generation of neutrino oscillation experiments sees $\nu_\mu \to
\nu_e$, then CP violation may be able to be probed by a conventional 
neutrino experiment.  
In order for CP violation to be seen, $\delta$
must be large or matter effects must be measured, and the experiments proposed
require extremely long run times (see table above) and depend on high 
background rejection in both neutrino and antineutrino running.  In this
scenario, the reach of a neutrino factory will be purely statistics 
limited, and precision measurements of all of the oscillation parameters
would be achievable.  

If KAMLAND does see neutrino oscillations but the next generation of neutrino 
experiments does not see evidence for $\sin^2 2\theta_{13}$ being non-zero, 
then CP violation and the sign of matter effects 
would only be accessible at a neutrino factory, as long as 
$\sin^2 2\theta_{13}$ 
was larger than a few $10^{-4}$.  For values smaller than that, $\nu_e$ to 
$\nu_\mu$ might still be seen, but it would be due to sub-leading oscillations,
or the solar mass scale.  
 
If KAMLAND does not see neutrino oscillations, then CP violation measurements
are not accessible at either a neutrino factory or a superbeam, and the 
physics we can hope to understand from oscillation measurements is ``only'' 
the size of $sin^2 2\theta_{13}$ and the sign of the largest mass splitting.  
Again, if evidence for 
$sin^2 2\theta_{13}$ is seen at the next generation of neutrino 
experiments, a neutrino factory could provide much better precision on 
this angle, as well as a guaranteed measurement of the sign of the largest
mass splitting.  If $\theta_{13}$ is not seen at the next generation of 
experiments, then neutrino factories have extremely good reach in this 
angle, since the subleading contributions due to the solar neutrino 
oscillations would not contribute significantly to the probability.  

It is clear that regardless of the outcomes of the next 
generation of neutrino experiments, a neutrino factory would 
provide a laboratory to extend our understanding of the lepton
mixing sector.  The fact that both the $\nu$ and $\bar\nu$ rates are 
so high, and the backgrounds in both beams are so low
allow huge leaps in measurement precision.  

\section{Paving the Road to a Neutrino Factory and Muon Collider} 

While the lead time to completing the R\& D for a neutrino factory 
is long, the machine itself can be constructed in stages to provide 
important physics opportunities at each step along the way 
\cite{stages}.  Because we do not know where the next big discovery
will lie, it is important to pursue the physics accessible at 
each of these stages.  The stages themselves can be described 
(simplifying) as follows:
\begin{enumerate} 
\item Upgraded Proton Source
\item Intense Muon Source: 200\,MeV 
\item Intense Muon Source: 2-3\,GeV  
\item 20-50\,GeV Muon Storage Ring
\item Muon Collider:  from a Higgs factory to the energy frontier
\end{enumerate} 

In the first part of this document we described the oscillation 
physics accessible with the neutrino beams which can come from 
this facility.  In the remainder of this document we discuss briefly 
the wealth of other measurements which can be made at the beamlines 
listed above.  

Aside from just the physics that we know can be done at these 
new facilities, it is important that R\& D be pursued for a neutrino 
factory for the following reason:  the neutrino factory itself came 
only as a byproduct of people trying to understand how 
to use muons to get to the energy frontier.  By exploring all the 
avenues we can for new experiments, we are opening the door for 
still more unforeseen techniques which may prove to be themselves 
landmark experiments in physics areas which we have yet to uncover.  

\section{Intense Muon Source Physics} 
\subsection{Overview}

An intense muon source, such as that provided for the front end of a
neutrino factory/muon collider, could yield significant improvements
in our exploration of muon physics. The expected intensity of such a
source is $10^{13}-10^{14}~\mu^{\pm}$/s, {\it i.e.}, five or six
orders of magnitude higher than that presently available.  Examples of
particle physics programs which might be pursued with intense muon
sources are (1) muon lepton flavor violation (LFV) and (2) muon
moments such as the anomalous magnetic moment ($g-2$) and the electric
dipole moment (edm) of the muon.  The former programs can best be done
with low-energy muons (mostly stopped muons), while the latter could
be carried out employing in-flight muons in a ring.

At present it appears that these muon physics programs would benefit
significantly from the staging accelerator scenario of a neutrino
factory. The coupling between the physics programs and staging is
illustrated in Table \ref{tb:stage}.  Stage I, a proton driver with
1--4 MW beam power, would yield significant improvements in the LFV
and muon moment experiments.  Stage II with a 200\,MeV cooled muon beam
would match well to an improved muon edm experiment.  Because the beam
repetition rate of such a source is low, however, new ideas on how to
handle high instantaneous rates would be necessary to utilize Stage II
for LFV.  With the 3\,GeV cooled muon beams of Stage III, the new
generation muon ($g-2$) experiment could be realized.

\begin{table}[h!]
\caption{muon physics programs in the accelerator staging approach.}
\label{tb:stage}
\bigskip
\begin{tabular}{|l|l|l|}\hline
Stage & Accelerator Component & Physics Programs \cr\hline\hline
I & high intense proton driver& LFV, muon edm, muon g-2 \cr\hline
II & 200 MeV cooled muon beam & muon edm, (LFV) \cr\hline
III & 3 GeV cooled muon beam & muon g-2 \cr\hline
\end{tabular}
\end{table}

\subsection{Muon Lepton Flavor Violation}

\subsubsection{Physics Motivation}

In the Standard Model (SM), LFV in charged lepton processes is
suppressed even with non-zero neutrino masses. However, in extensions
of the minimal SM LFV could occur from various sources
\cite{kuno01}. Important LFV processes involving muons are
$\mu^{+}\rightarrow e^{+}\gamma$, $\mu^{-}$-$e^{-}$ conversion in a
muonic atom ($\mu^{-} + N \rightarrow e^{-} + N$), $\mu^{+}
\rightarrow e^{+}e^{+}e^{-}$ and so on.

\begin{figure}
\includegraphics[width=\linewidth]{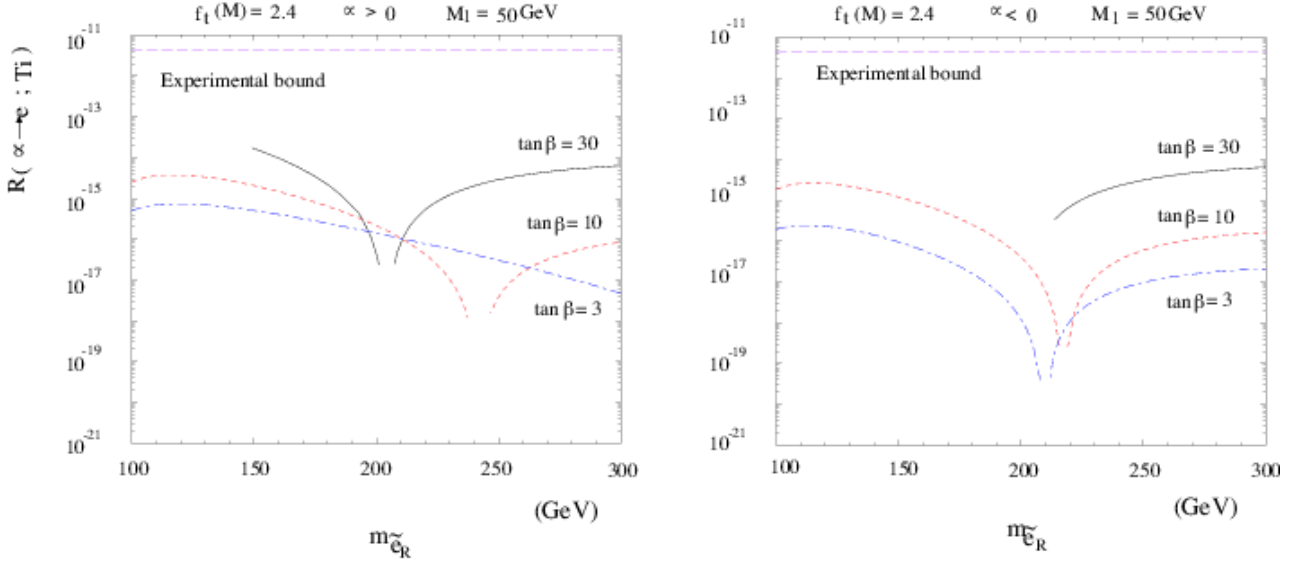} 
\caption{Prediction of $\mu^{-}-e^{-}$ conversion in SUSY SU(5).}
\label{fg:kuno-fig1}
\end{figure}

Recently, considerable interest in LFV has arisen based on
supersymmetric (SUSY) extensions to the SM, in particular
supersymmetric grand unified theories (SUSY-GUT).  In many models of
SUSY-GUT, LFV can be naturally introduced. For instance, in
supergravity-mediated SUSY models, radiative corrections in the
renormalization group evolution from the GUT scale to the weak-energy
scale lead to finite mixing in the slepton mass matrix, even when it
is assumed to be diagonal at the Planck scale.  Recently, Barbieri and
Hall \cite{barb94} found that the slepton mixing thus generated is
very large owing to the surprisingly large top quark Yukawa coupling.
Through loop diagrams $\mu \rightarrow e$ transitions then occur due
to this slepton mixing. The predicted branching ratio ranges between 
the current bounds and a few orders
of magnitude smaller \cite{strumia}, and could be
experimentally measurable.  The predicted branching ratio of
$\mu^{-}$-$e^{-}$ conversion in a muonic atom in SUSY SU(5) is shown in
figure \ref{fg:kuno-fig1}.

\begin{figure}[h!]
\includegraphics[width=\linewidth]{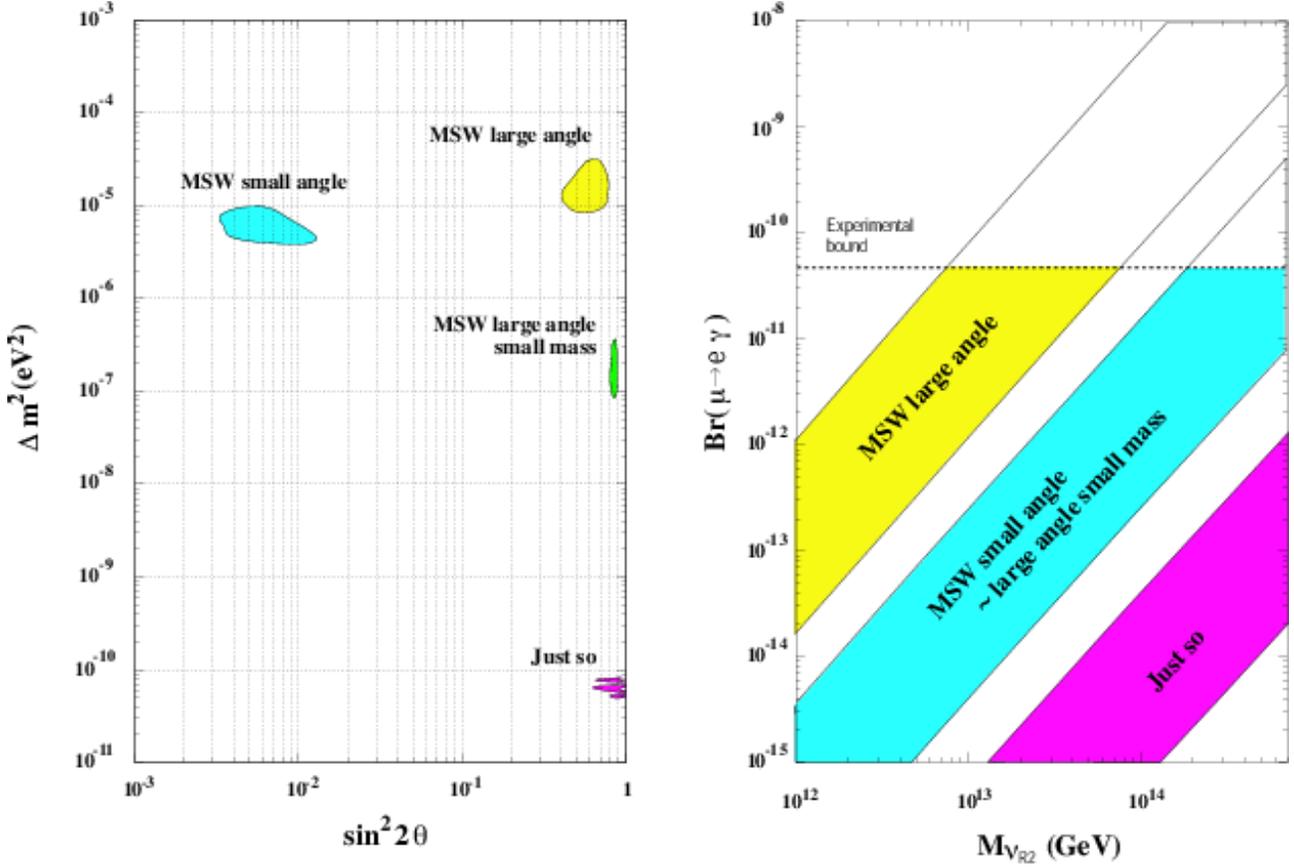}%
\caption{Left:  The allowed region in parameter spece for solar neutrino 
oscillations.  The regions in yellow, blue, and 
red correspond to the LMA, SMA, and VAC solutions, respectively.
Right: Prediction of the $\mu\to e \gamma$ branching ratio 
in SUSY models with heavy right-handed neutrinos. }
\label{susy-nu}
\end{figure}

Furthermore, the existence of massive neutrinos and their mixing, as
suggested by the recent solar and atmospheric neutrino measurements,
might allow additional LFV contributions in the SUSY-GUT models. Such
models includes a heavy right-handed majorana neutrino of
$10^{14}-10^{15}$ GeV/$c^2$ with $\nu_{\mu}-\nu_{\tau}$ mixing of
$\sin^2(2\theta_{32})\sim 1$. The predicted branching ratio for 
$\mu\to e \gamma$ is shown in figure \ref{susy-nu}.

CP violation in lepton flavor violation has been pointed out to be
important to study the Majorana CP phase of the heavy neutrino in the
see-saw model in a class of SUSY models \cite{ellis}. It could be
studied by T-odd correlation in $\mu^{+} \rightarrow e^{+}e^{+}e^{-}$
decay and the muon edm.

\subsubsection{Experimental Prospects}

Most of the LFV experiments could use stopped muons. There are the three 
such processes to study: (a) $\mu^{-}+N \rightarrow e^{-}+N$, 
(b) $\mu^{+} \rightarrow e^{+}\gamma$, and 
(c) $\mu^{+} \rightarrow e^{+}e^{+}e^{-}$.  The latter two
require a continuous beam in order to minimize accidental backgrounds. 
For a proton driver such as suggested for Stage I to be a source for these 
experiments, slow beam extraction with a high duty factor would be needed. 
On the other hand since process (a) is based on single
particle detection it does not suffer from accidental backgrounds, and 
therefore is the best suited for high rate muon beams. 

In what follows we discuss several experiments presented 
at the Snowmass conference.

\subsubsection{$\mu^{+}\rightarrow e^{+}\gamma$}

A detector for $\mu^{+}\rightarrow e^{+}\gamma$ has to have good
energy and position resolutions as well as timing resolution for
$e^{+}$ and $\gamma$. A new experiment to aim at a sensitivity of
$10^{-14}$ at PSI is being prepared, with a xenon photon
calorimeter. To go beyond, a detector improvement is necessary before
an increase of a muon beam intensity.

\subsubsection{$\mu^{-}-e^{-}$ conversion in a muonic atom}

\begin{figure}
\includegraphics[width=10cm]{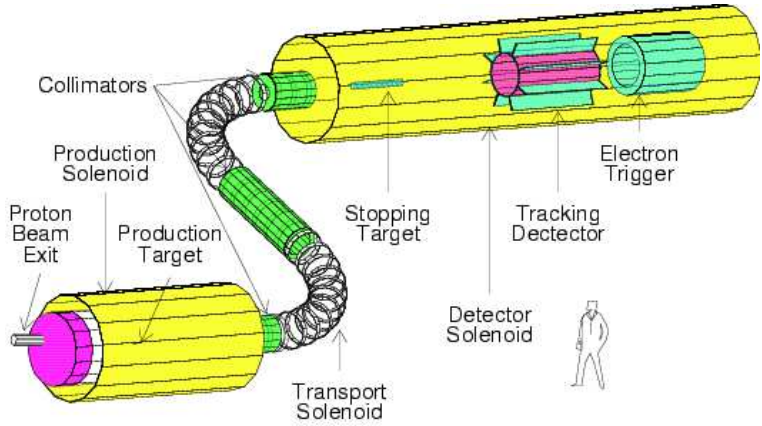}%
\caption{Schematic layout of the MECO detector.}
\label{fg:meco}
\end{figure}

The current upper limit of $B(\mu^{-}+$Ti$\rightarrow e^{-}+$Ti$)<6\times
10^{-13}$ comes from the SINDRUM-II experiment at PSI. A new
experiment, E940 (MECO) is being prepared at BNL-AGS \cite{meco}. It
aims to search for $\mu^{-}+$Al$ \rightarrow e^{-}+$Al at a sensitivity
better than $10^{-16}$.  The experimental setup is shown in
figure \ref{fg:meco}.  A pulsed proton beam of about 600\,kHz with pulse
width of 50\,nsec is used to minimize beam-associated backgrounds. The
muon beam rate of $10^{-11} \mu^{-}$ per second stopping in the target
is expected with 50\,kW proton target energy deposit. The single rate
of detector chambers is as high as 500\,kHz.

What would be possible upgrades once an intense proton driver and the
front end of neutrino factory are available?  The MECO detector is
optimized for the current BNL-AGS beam rate. If an intense proton
driver (of 1-4\,MW) were available, further detector optimization would
be needed to reduce the singles rates, but such optimization is
feasible.  At stage II it would be advantageous to use a cooled muon
beam with a smaller energy spread and smaller beam size. However, the
low repetition rate (a few tens of Hz) would be a drawback.  Perhaps a
smart detector system might be developed to handle the high
instantaneous rates.

In Japan, a dedicated facility called  PRISM (=Phase Rotated Intense Slow 
Muons) is being considered for the JHF. It employs 
phase rotation to reduce the beam energy spread, and has a long flight 
path in a fixed-field-alternating-gradient synchrotron ring (FFAG) to remove 
the surviving pion contamination in the muon beam. 

Whatever the nature of an intense muon source, a next-generation experiment 
could aim at a sensitivity of $10^{-18}$.

\subsection{Muon Moments}

\subsubsection{Muon $g-2$ Magnetic Moment}

The recent observation of a 2.6 $\sigma$ deviation from the Standard Model 
prediction of the anomalous magnetic moment of the muon ($g-2$) is a dramatic 
and exciting result for particle physics.\cite{g2}.   It might well be
an indicator of physics beyond the Standard Model. If such new physics
originates from SUSY, it may also give rise to muon LFV; non-Standard Model
$g-2$ is sensitive to the diagonal matrix elements of the slepton mass matrix, 
while LFV senses the off-diagonal matrix elements. 

Now that $g-2$ has been measured to the parts per million (ppm) level an
improved understanding of systematic uncertainties in this type of
measurement is available.  It is anticipated that in a new experiment 
the the basic technique would not change: a weak focusing storage ring 
operating at 3.1 GeV, with electrostatic quadrupoles providing vertical
focusing, and electrons from decaying stored muons being observed.  
A more uniform magnetic field, smaller storage aperture, and better 
magnetometer and beam inflector, along with the reduced phase space and more 
intense beam provided by a Stage III cooled muon beam, could provide a 
significantly improved measurement.  Increased polarization with a 
small sacrifice in intensity is an added benefit.  All of the above 
improvements would serve to provide a more precise measurement with
smaller systematic uncertainty.

On the theoretical side, more precise experimental measurements of 
hadron production from $e^+ e^-$ collisions at low energies and $\tau$ lepton
decays coupled with better lattice calculations could reduce uncertainties 
in $g-2$ due to the hadronic contribution.

\subsubsection{Muon Electric Dipole Moment}

There are already stringent limits on the edm of 
first generation particles (electron and neutron). It is very important to 
do a sensitive search for the edm of a second generation particle. The 
Feynman diagram is the same as for the muon anomalous magnetic moment, but
with a CP violating phase. The Standard Model prediction for all edm's is 
unmeasurably small. Therefore, any measured value indicates new physics 
such as supersymmetry. Several
models predict values for the muon edm at the few 
$\times 10^{-23}$ e-cm level, 
consistent with electron edm limits and the muon $g-2$ value. 

The last measurement of the muon edm gave ($3.7 \pm 3.4$) $10^{-19}$ e-cm.
There is an LOI for
a dedicated muon electric dipole moment experiment at the BNL-AGS. 
It plans to use the $g-2$ storage ring, but 
its field would be set below the magic 
momentum used by the $g-2$ experiment (3.1 GeV/$c$). Presently, 
the optimum momentum is believed to be between 0.2 and 0.5 GeV/c.  

Such an experiment is envisioned to proceed in three stages:

(1) Needed statistics: $NP^2 = 10^{12}$, where $N$ is the number of muon 
decays accumulated
and $P$ is the polarization.  
The existing $g-2$ ring would
be changed to weak magnetic focusing (instead of electrostatic). The level 
of sensitivity to the edm would be $\sim 10^{-22}$ e-cm after about 400 hours 
of physics running, and the systematic
uncertainties could be measured with an accuracy of 10$^{-24}$ e-cm. This 
stage provides very important information about the real problems in such a 
measurement.

(2) $NP^2 = 10^{14}$. The $g-2$ storage ring would be further modified to 
strong focusing. The level of
sensitivity to the edm would then be $\sim 10^{-23}$ e-cm after about 4000 
hours of physics running.  A Proton Driver would reduce the 4000 hour to 
500 hours.

(3) $NP^2 = 10^{16}$. This probably requires stage II of the front end of 
neutrino factory \cite{edm}.  At this stage, 
$NP^2 = 10^{16}$ can be accumulated
in  about one year of physics running.  Almost certainly, a new storage ring
would  be necessary, optimized for this measurement. Note that
unlike the $g-2$ experiment, it is not necessary to keep the homogeneity 
of the magnetic field to the $10^{-7}$ level, but
only to what is usually required for storage rings (about three orders of 
magnitude worse). The
required beam specification is (a) muon beam momentum:  0.2-0.5 GeV/c, 
(b) muon beam intensity: $10^{11}$/sec, (c) $NP^{2}$: $10^{16}$, 
(d) dp/p: ~1\%, (e) angular
divergence: $\approx$10\,mrad, (f) beam size: 
$\approx$100\,mm, (g) bunch duration: $<30$\,ns, 
(h) time between bunches: $20$\,ms, (i) polarization: as large as practical 
(a modest 16\% in the neutrino factory design report).

\subsection{Applications}

Once an intense muon source is available, other applications 
can be considered. They are, for instance, muon catalyzed fusion, and life 
science studies by $\mu$SR (muon spin rotation). For the former, a small 
size target of D-T mixture could be exposed to extreme conditions with a high 
intensity $\mu^{-}$ beam. For the latter, the use of a smaller sample with a
small phase space, highly polarized muon beam could provide 
critical next-generation experiments. Besides these two, various applications 
to materials science are also envisaged.

\section{Neutrino Scattering Physics} 

\subsection{Introduction}

There is still much to be learned {\bf from} neutrinos as well as much to
be learned {\bf about} neutrinos.  Although neutrino oscillation experiments
certainly will drive the construction of new neutrino beamlines, these new
very intense beamlines also allow us to continue an active research program
at a detector located close to the production target.  At such a near
detector, associated with a superbeam or neutrino factory, the event rates
will be much higher than at the previous generation of neutrino beam
facilities allowing the use of much lighter targets and avoiding the large 
and unknown nuclear effects which complicate the interpretation of current
neutrino scattering experimental results.

\subsection{Low Energy Neutrino Physics}

There are many interesting topics which can be studied as part of a
low energy neutrino program at a superbeam
facility.  Some of these are only possible with the high intensities
expected there.  Physics topics using low energy, high
intensity neutrino beams explored at Snowmass include:

\begin{itemize}
\item  $\nu_\mu e^-$ elastic scattering at low-$Q^2$; the neutrino magnetic
 moment.
\item Quasi-elastic scattering; the strange-spin of the nucleon, $\Delta s$
\item Lepton number violating processes (non-oscillations) 
\end{itemize}

\subsubsection{Neutrino-electron elastic scattering--
Search for nonzero neutrino magnetic moment}

The recent discoveries in the neutrino sector in the Standard Model
have opened a new frontier in high energy physics.  Understanding
neutrinos and how they interact is crucial to continuing to verify the
Standard Model and look for physics beyond the Standard Model.  Searches
for electromagnetic properties of neutrinos, such as a non-zero
neutrino magnetic moment, can set limits on beyond Standard Model
physics in the neutrino sector and in other sectors as well as
addressing a number of important astrophysical limits.

Neutrino magnetic moments can arise through a variety of beyond the 
Standard Model mechanisms.  In the minimally extended Standard Model
massive Dirac neutrinos of mass $m_\nu$ can have a non-zero neutrino
magnetic moment of the size $\mu_\nu = \frac{3eG_F}{8 \sqrt{2}\pi^2}
m_\nu \sim 3 \times 10^{-19} \mu_B \frac{m_\nu}{1 eV}$ arising from
one loop radiative corrections in diagrams with W-boson exchange.
Extensions to the Standard Model such as the supersymmetric left
right model and models including large extra dimensions predict
neutrino magnetic moments ranging up to $10^{-11} \mu_B$
~\cite{frank,EDs}.  A non-zero neutrino magnetic moment would also
have important implications in cosmology in the development of stellar
models.  Astrophysical limits such as plasmon decay rates from
horizontal branching stars and neutrino energy loss rate from
supernovae allow a neutrino magnetic moment as large as
$10^{-11}\mu_B$~\cite{astro1,astro2}.

A non-zero neutrino magnetic moment can give rise to an
electromagnetic contribution to neutral current neutrino scattering.
This is most easily measured using neutrino-electron elastic
scattering.  Present experimental limits for the muon neutrino
magnetic moment come from the LSND experiment which sets an upper
limit of $\mu_{\nu_\mu} < 6.8 \times 10^{-10} \mu_B$ \cite{LSND2} by
measuring the total $\nu$-$e$ elastic scattering cross section.

At low $y=\frac{E_e}{E_\nu}$ the electromagnetic contribution to the
$\nu-e$ cross section increases rapidly while the Standard Model
contribution increases only gradually.  This shape dependence can be
used to look for a signal and possibly greatly extend our
sensitivity to a non-zero neutrino magnetic moment into the region
where beyond the Standard Model theories predict and astrophysical limits
allow non-zero neutrino magnetic moments.  A high intensity, low
energy ($E_\nu \sim 1~GeV$) neutrino beam such as that available at a
proton driver or superbeam facility can make this measurement
possible.

The sensitivity to a neutrino magnetic moment in a stage I neutrino 
beam has been calculated, 
assuming a MiniBooNE-size detector located 100\,m from the
neutrino source (assuming a proton driver upgrade to the MiniBooNE
beamline).  For a reasonable range in detectable recoil electron
energy thresholds, 
the statistical sensitivity on $\mu_{\nu_\mu}$ is in the few times 
$10^{-11}$ range.  

If a $10\%$ systematic error due to the flux is included, the dominant
systematic error in such experiments, the sensitivity to $\mu_{\nu}$
becomes $2.2 \times 10^{-10} \mu_B$ and the experiment becomes
systematics-limited.  A better method for determining any
electromagnetic contribution to the cross section is to take advantage
of the shape dependence of the differential cross section on an
electromagnetic component of the interaction.  This method does not
require precise knowledge of the flux.  Work on determination of sensitivity
to $\mu_{\nu}$ using this method continues.

Non-traditional methods to search for neutrino magnetic moment using
higher energy neutrino beams are 
discussed in reference~\cite{Neutrinofactory}, but their feasibility 
has not yet been
demonstrated.

\subsubsection{A Future Experiment to Measure $\Delta s$ with a high
intensity neutrino source}

A topic of large and continuing interest in nuclear and particle
physics is the role of strange quarks in the properties of the
nucleon. Neutrino nucleon elastic scattering is sensitive to an
isoscalar contribution to the nucleon spin via the $\nu p$ axial
coupling.  This is presumably the same contribution responsible for
the surprising results from the EMC experiment (and subsequent) that
show a violation of the Ellis-Jaffe sum rule.

A measurement of $\nu p$ elastic scattering in the kinematic range
accessible to the MiniBooNE experiment ($0.1<Q^2<1.0~GeV^2/c^2$)
with sufficient reduction of systematic errors allows for a precise
extraction of the $\nu p$ axial form factor ($G_A$) and the 
contribution of strange quarks to the spin of the nucleon, $\Delta s$.

By measuring $\nu p$ elastic (neutral current) scattering and
comparing to $\nu n$ quasi-elastic (charged current) scattering, a
sensitive measurement of the ``strange'' part of $G_A$, $G_s$, 
may be obtained with little
systematic error due to the uncertainty in the neutrino flux.  
(Note in the limit $Q^2=0$, $G_s(0)=\Delta s$.)

In addition, if it is possible to measure $\nu n$ elastic scattering
with sufficient
precision, and the neutral and 
charged current cross sections with antineutrinos, this 
data set would allow a very robust extraction of $\Delta s$ 
along with the axial form factor mass, $M_A$. 

The detector for this measurement would require high segmentation
for tracking to separate the final state particles.  It would also
need a moderate level of particle identification capability to 
distinguish the possible background reactions.

A high intensity neutrino source such as a proton driver beam or a
superbeam would benefit this experiment by increasing the events rates
tremendously.  This could allow a smaller detector with better
segmentation and tighter cuts on the event sample to better understand
systematic errors.

\subsection{Medium-to-High Energy Neutrino Physics} 

A superbeam or neutrino factory providing intense neutrino beams in
the 2 - 20 GeV range will enable a study of the surprisingly still poorly 
understood region of neutrino resonance production, the transition from
resonance to DIS and certain kinematic regions of DIS.  To study these
mechanisms by scattering neutrinos off a 
light target will allow us to finally answer many pending
questions.

\subsubsection{Neutrino Deeply Inelastic Scattering}

Neutrino-nucleon experiments offer a rich source of information about the
quark structure of the proton.~\cite{nuRev} Neutrino-nucleon deeply inelastic
scattering (DIS) is arguably the most direct measurement of the proton
structure functions.  However, at low-$Q^2$, and high-x disentangling
perturbative effects from nuclear effects and higher-twist effects becomes
extremely difficult.  The neutrino DIS events with $Q^2 < 1.25$ GeV$^2$ and
for $x < 0.1$ as well as $x > 0.6$ are largely unused because of these effects.

With the high statistics foreseen at a superbeam or a neutrino factory,
allowing the use of light targets, as well as the special attention to
minimizing neutrino beam systematics necessary for neutrino oscillation
experiments, it should be possible for the first time to determine the
separate structure functions $2F_1^{\nu N}(x,Q^2)$,$ 2F_1^{\bar \nu
N}(x,Q^2)$, $F_3^{\nu N}(x,Q^2)$ and $F_3^{\bar \nu N}(x,Q^2)$ where N is an
isoscalar target. In leading order QCD (used for illustrative purposes) these
four structure functions are related to the parton distribution functions by:

\begin{eqnarray}
2F_1^{\nu N}(x,Q^2) &=& u(x) + d(x) + s(x) + \bar u(x) + \bar d(x) + \bar
c(x)
\nonumber \\
2F_1^{\bar \nu N}(x,Q^2) &=& u(x) + d(x) + c(x) + \bar u(x) + \bar d(x) +
\bar s(x)
\nonumber \\
xF_3^{\nu N}(x,Q^2) &=& u(x) + d(x) + s(x) - \bar u(x) - \bar d(x) - \bar
c(x)
\nonumber \\
xF_3^{\bar \nu N}(x,Q^2) &=& u(x) + d(x) + c(x) - \bar u(x) - \bar d(x) -
\bar s(x) .
\nonumber
\end{eqnarray}

Note that taking differences and sums of these structure functions would then
allow extraction of individual parton distribution functions in a given
$x,Q^2$ bin:

\begin{eqnarray}
2F_1^{\nu N} - 2F_1^{\bar \nu N} &=& [s(x) - \bar s(x)] + [\bar c(x) -
c(x)]
\nonumber \\
2F_1^{\nu N} - xF_3^{\nu N} &=& 2[\bar u(x) + \bar d(x) + \bar c(x)]
\nonumber \\
2F_1^{\bar \nu N} - xF_3^{\bar \nu N} &=& 2[\bar u(x) + \bar d(x) + \bar
s(x)]
\nonumber \\
xF_3^{\nu N} - xF_3^{\bar \nu N} &=& [\bar s(x) + s(x)] - [\bar c(x) +
c(x)] .
\nonumber
\end{eqnarray}

As we increase the order of QCD and allow gluons into consideration we need to
bring in global fitting techniques into the extraction of the parton
distribution functions.  However, if the statistical and systematic errors can
be kept manageable, the ability to isolate individual parton distribution
functions will be dramatically increased by measuring the full set of separate
$\nu$ and $\bar \nu$ structure functions.

\subsubsection{High-x Parton Distribution Functions}

There is considerable interesting physics in the region of high $x$. This
region can be described as the ``bridge'' between perturbative QCD and
non-perturbative QCD, a bridge that Lattice Gauge Theory is trying to
construct. This is a region that requires much additional study with both
electroproduction and the weak current of neutrino nucleon interactions.

In global fits of experimental data to extract the parton distribution
functions, the functions -- even the gluon distribution 
-- are fairly well known from
very small $x$ 
up to $x$ of around 0.5.  Above this value there is very little data
and, in particular, all neutrino data in the region is on heavy nuclear
targets and subject to strong nuclear effect which have never been
measured. A high statistics neutrino/antineutrino exposure in $H_{2}$ and
$D_{2}$ provides the most direct way of studying this rich region of phase
space.

   The uncertainties at high $x$ in current nucleon parton distribution
functions are of two types: the ratio of the light quark PDF's, $d(x)/u(x)$,
as $x\to1$ and the role of leading power corrections (higher twist) in the
extraction of the high $x$ behavior of the quarks.  These higher twist (or
power suppressed) corrections represent a long-standing hurdle to making
accurate theoretical predictions for structure function data over the full
kinematic range.  Higher twist corrections should not simply be avoided;
accurate characterization of higher twist corrections provides new
information on parton-parton correlations within the nucleus.

The kinematic limits where considerations of higher twist contributions
become important are 1) at high-x, and 2) at low $Q^2$ where terms of order
$\Lambda^2/Q^2$ become significant.  In the high-x region, the limiting
factor is primarily statistics.  In the low $Q^2$ region the statistics are
generally adequate, but if the data is taken on heavy targets the higher
twist effects are entangled with nuclear effects.

Consequently, the ideal testing ground would be to have high statistics
measurements on a light target. This would allow systematic separation of
the higher twist effects from the nuclear effects, and better allow us to
learn about both in the process.

Another challenge of high-x physics is the long-standing problem in QCD --
the calculation of the cross section for the production of heavy quarks
both in hadroproduction and leptoproduction mode. In the case of the
b-quark, for example, there are large discrepancies between data and theory
both at the Tevatron and at HERA facilities.  Another unsettling aspect of
heavy quark production is the relatively large theoretical uncertainty
remaining in the calculations, despite the existence of next-to-leading
order calculations.

One degree of freedom that has not been fully studied or exploited in
this area is the issue of an ``intrinsic" heavy quark component. While
the question of intrinsic heavy quarks has been discussed in the
literature for many years, it still remains unresolved: a definitive
experiment is needed!  A particularly incisive test of this theory
would be to make precise measurements of heavy quark production in the
threshold region.  In this kinematic regime, the usual ``perturbative"
heavy quark component arising from gluon splitting ($g\rightarrow Q
\bar{Q}$) is comparatively small; therefore a measurement in this
region has more discriminating power to confirm or refute the ``intrinsic"
heavy quark component once and for all.  Experiments studying 
heavy quark production would occur at the 20-50GeV neutrino factory (Stage IV).

\subsubsection{A Detector for Future Neutrino Scattering Experiments}

A number of proposed projects that use a high-luminosity proton driver to
generate a neutrino beam for oscillation physics could add a light mass
detector for the purpose of studying low-$Q^2$ neutrino DIS.

Conceptually, the type of detector that would be required would be one
which has excellent hadronic and muon energy and angle resolution as
well as particle identification.  Due to the considerable resonance
contribution to the neutrino cross section at low energies, particle
id is required for identification of resonance events.

Examples of low-mass detectors which have the required particle id are
liquid-He or liquid-H$_2$ time projection chambers.  The information
acquired from these chambers is very similar to that of bubble
chambers.  These detectors feature excellent particle identification
and good momentum and energy resolution.  

\subsection{Neutrino Flavor Violation Physics --non-oscillations} 

A lepton flavor violation program involving {\em neutrinos} rather
than muons would nicely complement the muon physics and neutrino
oscillation physics programs at a staged neutrino factory. Several
signatures could be looked for in a neutrino experiment with a short
enough baseline so that oscillations can be neglected. For example, in
the case of neutrino production from $\mu^+$ decays, the detection of
wrong sign muons, positrons, taus of both signs would all signal new
physics in the decay or in the neutrino interaction in the
detector. In particular, $\mu^-$ could arise from the standard CC
interaction of muon neutrinos from the flavor violating decay
$\mu^+\rightarrow e^+ \nu_\mu\bar\nu_l$ ($l=e,\mu,\tau$), as well as
form the non-standard $\bar\nu_\mu$ interaction $\bar\nu_\mu \;
e^-\rightarrow \bar\nu_l\;\mu^-$ or the non-standard $\nu_e$
interactions $\nu_e\; d\rightarrow \mu^-\; u$, $\nu_e\; e^-\rightarrow
\nu_e\; \mu^-$. Analogously, positrons could be produced in the
standard CC interactions of electron antineutrinos from
$\mu^+\rightarrow e^+\nu_l\bar\nu_e$ or from the process
$\bar\nu_\mu\; u\rightarrow e^+ d$ in the detector. In each case,
observation would immediately point to new physics well beyond the
implications of the Standard Model.

Particularly interesting from the experimental point of view is the
wrong sign muon signature.  By using a $L=100\,{\rm m}$
baseline, a muon energy of 2\,GeV (Stage III), and a 10 ton detector, 
the sensitivity on the
branching ratios ${\rm BR}(\mu^+\rightarrow e^+\nu_\mu X)$ and ${\rm
BR}(\mu^-\rightarrow e^- \bar\nu_\mu X)$ is improved by two orders
of magnitude below current values, 
which only extend to the $10^{-2}$ level ~\cite{hep-ph/0010308} .

In the case of higher neutrino energies, it is possible for neutrino
detectors to be sensitive to additional lepton family violating
channels. Specifically, for neutrino energies above 10.7 GeV (stage IV), 
one can
be sensitive to the reaction $\bar \nu_\mu e^- \rightarrow \mu^- \bar
\nu_e$.  Such an interaction is both sensitive to new physics, such as
left-right symmetry and dileptons, and has the very clean wrong sign
muon signature.  In addition, because the target is an electron, the
muon emanating from the reaction will have very small opening angle
($p_t^2 \le m_e E_\nu /2$), which can be used as an additional handle
to distinguish events from Standard Model and non-Standard Model
background.  This possibility has also been studied
in~\cite{ex:nutev-lnv}. The same detectors as outlined in
Silicon CCD or liquid methane TPC detectors can be used which 
provide excellent charge,
vertex, and angular resolution ~\cite{bruce}. 
Projected sensitivities at Stage IV would improve over current 
limits by 3-4 orders of magnitude and, if backgrounds can 
remain under control, reach the $10^{-6}$ or $10^{-7}$ level.  

The scale of new physics that can be probed with such a sensitivity
depends on the specific model one considers. However, the cleanliness
of the experimental signature and its complementarity to neutrino
oscillation experiments makes lepton flavor violation searches an
attractive feature of a stage IV neutrino factory program.  

\section{Muon Collider Physics}

Although muon colliders were discussed at Snowmass '96 as 
energy frontier machines, much has been learned in the meantime about 
what physics could be achieved on the way to such a device.  The 
experiments described in this document have so far borne little resemblance
to the experiments that were proposed at the previous Snowmass workshop.  
In this last section we describe
the motivation for using a muon collider as a Higgs factory (i.e., still 
much lower in energy that was originally proposed), what physics it 
could provide, and what some of the detector concerns are for this 
experiment.  

\subsection{Physics Issues}

At a muon collider~\cite{muon_coll} the Higgs boson would be 
produced through the
$s$-channel, so the production cross section is thousands of times
larger than the cross section at an $s$-channel $e^+ e^-$ collider. 
Because a muon beam energy spread as small as $\sim
10^{-5}$ may be possible, there is a possibility of measuring \mH\ to a few
hundred keV and a direct measurement of the width to about 1 MeV.  If
only one light Higgs boson were observed, it would be crucial to 
measure its properties to infer
whether it is a Standard Model or supersymmetric Higgs. The $CP$
properties of the Higgs bosons can be measured through asymmetries
with transversely polarized $\mu^+$ and $\mu^-$ beams~\cite{bbgh}. In the
case of heavy MSSM Higgs bosons, the large coupling to $\mu^+ \mu^-$
may be necessary for their direct observation.

Although the Higgs boson mass must be known to a few per cent \cite{bbgh}  
before knowing at what energy 
to build a muon collider as a Higgs Factory, it is believed that that 
mass is low.  Once a beam of 50\,GeV muons can be collected and stored
in a ring to do neutrino experiments, the remaining task to get to a 
Higgs factory would be mostly an issue of beam cooling, since the 
center of mass energy of two 50\,GeV muon beams is expected to be relatively 
close to the Higgs mass.  

In order to measure the width of a narrow (2-3 MeV) Higgs boson of mass
(~120GeV), one needs to have beam energy spread and stability of order
$~10^{-5}$ and also to measure the energy of the bunches to $10^{-6}$. 
The latter
measurement is feasible using $g-2$ spin precession of the muons by measuring
\cite{gtworaja} the energy spectrum of the decay electrons turn by turn.

Indirect information about the mass of the Standard Model Higgs boson
can be obtained from fits to the precision electroweak data taken at
the $Z^0$ resonance at LEP and the SLC, and from neutrino-Nucleon 
Deep Inelastic scattering cross section 
measurements.  The Z-pole cross sections and asymmetries are
sensitive to the mass of the top quark \mt, the mass of the $W$
boson \mw, the QCD coupling constant $\alpha_s$.  Most electroweak 
observables are sensitive 
to the log of the mass of the Higgs boson \mH\ 
through radiative corrections. 
The electroweak data fit gives~\cite{charlton} 
\mH\ = 88$^{+53}_{-35}$ GeV and \mH\
$<$ 196 GeV at 95\% C.L.   

Although electroweak fits suggest a light Higgs, searches thus far
have produced only lower bounds on the mass.  
At LEP the SM Higgs boson is expected to be produced mainly through
the Higgs-strahlung process $e^+ e^- \rightarrow H^0 Z^0$, with
contributions from the $WW$ fusion channel below 10\%. 
%The likelihood
%analysis of the combined data from the four LEP experiments shows a
%preference for a Higgs boson with a mass of 115.6 GeV. At this mass,
%the probability for the Standard Model background to generate the 
%observed effect is
%3.4\%. 
The lower bound coming from a combined analysis of the four LEP 
experiments is $\mH > 114.1$  GeV at 95\% C.L. (115.4 GeV expected).

%The LEP experiments requested additional running in 2001 to increase
%the significance of a possible signal to the level of $\sim 5 \sigma$,
%which would have conservatively required about six months total
%running time~\cite{lep_request}. However, the request was not
%approved. Now we will have
%to wait until $\sim$ 2007 to find out from the LHC experiments, or
%possibly the Fermilab Tevatron, whether there really is a Higgs boson
%at a mass of $\sim$ 115 GeV.

%\subsubsection{Muon Anomalous Magnetic Moment}%%

%Recent results~\cite{anom} from BNL E821 show a disagreement of the
%anomalous magnetic moment of the muon $a_\mu \equiv (g-2)/2$ with the
%Standard Model prediction by $2.6 \sigma$. The deviation is consistent
%with contributions from loops due to supersymmetric particles.

%Data equivalent to four times the amount analyzed have been recorded
%but not yet analyzed. Analysis of this larger data sample may be
%completed by the end of 2001.

\subsubsection{Implications for Supersymmetry}

The hints for a Higgs boson with a low mass and the
disagreement of $(g - 2)_\mu$ with the Standard Model expectation \cite{g2} 
are consistent with the following general scenario~\cite{bbgh}:

\begin{itemize}
\item
In the Minimal Supersymmetric Standard Model (MSSM), $\mh \sim 115$
GeV indicates a large value of $\tan \beta$. 
\item
The disagreement of $(g - 2)_\mu$ also indicates 
a large value of $\tan \beta$. 
\item
If the
disagreement of  $(g - 2)_\mu$ is explained by supersymmetry, then the
sign of the supersymmetry parameter $\mu$ is consistent with $b
\rightarrow s \gamma$. 
\item
In the decoupling limit, the lighter Higgs
boson $h^0$ has couplings like the Standard Model Higgs, but the heavier
Higgs bosons $H^0$, $A^0$ have non-Standard Model couplings: their coupling to
gauge bosons is greatly suppressed. 
\item
For larger values of $\tan \beta$,
there is a range of heavy Higgs boson masses for which discovery is
not possible at the LHC or an $e^+ e^-$
linear collider. 
\item
In the MSSM, the heavy Higgs bosons are largely
degenerate, especially in the 
decoupling limit. Very precise
center-of-mass energy resolution will be needed to separate them.
\end{itemize}

\subsection{Muon Collider Detectors}

\begin{figure}[bth!]
\includegraphics[width=0.5\linewidth]{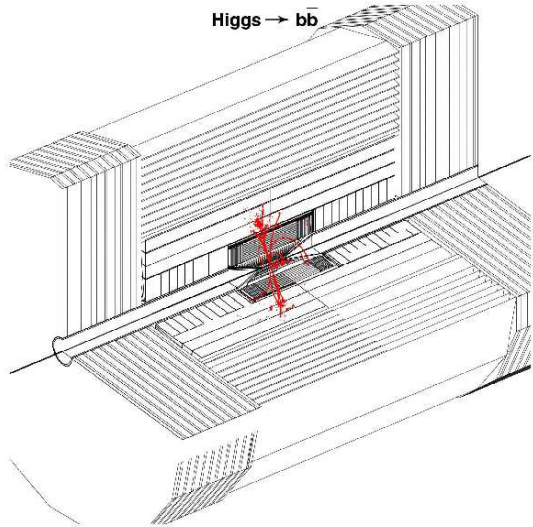} 
\caption[GEANT detector for a muon collider]
{Cut view of a potential detector in GEANT for the Higgs factory with
a Higgs$\rightarrow b\bar b$ event superimposed. No backgrounds are
shown. The tungsten cones on either side of the interaction region
mask out a $20^\circ$ area.
\label{geant}}
\end{figure}

Figure~\ref{geant} shows a trial
muon collider detector for a Higgs factory simulated in GEANT. The
background from muon decay sources has been extensively
studied~\cite{muon_coll}. At the Higgs factory,
the main sources of background are from photons generated by the
showering of muon decay electrons. At the higher energy colliders,
Bethe-Heitler muons produced in electron showers become a problem.
Work was done to optimize the shielding by using specially shaped
tungsten cones~\cite{muon_coll}. 
The background rates obtained were shown to be
similar to those predicted for the LHC experiments. It still needs to
be established whether pattern recognition is possible in the presence
of these backgrounds.

\section{Conclusions}
 
The recent discovery of neutrino oscillations is a profound discovery.
The US should strengthen its lepton flavor research
program by expediting construction of a high-intensity, conventional neutrino
beam ("superbeam") fed by a 1--4\,MW proton source.
 
A superbeam will probe the neutrino mixing angles and mass hierarchy, and may
discover leptonic CP violation.  The full program will require neutrino beams
at a number of energies, and massive detectors at a number of baselines.  
These
facilities will also support a rich program of other important 
physics, including
proton decay, particle astrophysics and charged lepton CP- and flavor- 
violating processes.

The ultimate laboratory for neutrino oscillation measurements is a neutrino 
factory, for which the 
superbeam facility serves as a strong foundation.  
The development of the additional needed technology for neutrino 
factories and muon colliders requires an ongoing vigorous 
R\&D effort in which the US should be a leading partner.   

% figures should be put into the text as floats.
% Use the graphicx package (distributed with LaTeX2e).
% See the LaTeX Graphics Companion by Michel Goosens, Sebastian Rahtz,
% \begin{figure}
% \caption{}
% \label{}
% \end{figure}

% tables follow here or maybe be put in the text
% \begin{table}
% \caption{}
% \label{}
% \end{tabular}
% \end{table}

%\bibliography{your bib file}
\begin{acknowledgments}
The authors are grateful to Ed Stoeffhaus for maintaining the 
E1 Working group web pages through their many iterations: 
before, during, and after the Snowmass Workshop.  
\end{acknowledgments}

\end{document}